\documentclass[twocolumn,showpacs,preprintnumbers,superscriptaddress,amsmath,amssymb,nofootinbib]{revtex4}
\usepackage{graphicx}% Include figure files
\usepackage{dcolumn}% Align table columns on decimal point
\usepackage{bm}% bold math

\usepackage[normalem]{ulem}  % \sout{old text} for strikeout
\usepackage[dvips]{color} % For blue in-text comments and additions

\renewcommand\sout{\bgroup \color{red} \ULdepth=-.5ex \ULset}

\newcommand{\Psfig}[2]{\includegraphics[width=#1]{#2}}
\newcommand{\PsfigII}[2]{\includegraphics[scale=#1]{#2}}

\newcommand{\SUN}[1]{\text{SU} ( #1 )}

\def\kev{\text{ keV}}
\def\mev{\text{ MeV}}

\def\trace{\text{tr}}

\begin{document}

\preprint{}

\title{Determination of compositeness of the $\bm{\Lambda (1405)}$
  resonance from its radiative decay}% Force line breaks with \\

\author{T.~Sekihara} 
\email{sekihara@post.kek.jp}
\affiliation{KEK Theory Center, Institute of Particle and Nuclear
  Studies, High Energy Accelerator Research Organization (KEK), 1-1,
  Oho, Tsukuba, Ibaraki 305-0801, Japan}

\author{S.~Kumano}
\affiliation{KEK Theory Center, Institute of Particle and Nuclear
  Studies, High Energy Accelerator Research Organization (KEK), 1-1,
  Oho, Tsukuba, Ibaraki 305-0801, Japan}
\affiliation{J-PARC Branch, KEK Theory Center,
  Institute of Particle and Nuclear Studies, 
  High Energy Accelerator Research Organization (KEK),
  203-1, Shirakata, Tokai, Ibaraki, 319-1106, Japan}

\date{\today}% It is always \today, today,

\begin{abstract}
  The radiative decay of $\Lambda (1405)$ is investigated from the
  viewpoint of compositeness, which corresponds to the amount of
  two-body states composing resonances as well as bound states.  For a
  $\bar{K}N (I=0)$ bound state without couplings to other channels, we
  establish a relation between the radiative decay width and the
  compositeness.  Especially the radiative decay width of the bound
  state is proportional to the compositeness.  Applying the
  formulation to $\Lambda (1405)$, we observe that the decay to
  $\Lambda \gamma$ is dominated by the $K^{-}p$ component inside
  $\Lambda (1405)$, because in this decay $\pi ^{+} \Sigma ^{-}$ and
  $\pi ^{-} \Sigma ^{+}$ strongly cancel each other and the $\pi
  \Sigma$ component can contribute to the $\Lambda \gamma$ decay only
  through the slight isospin breaking.  This means that the decay
  $\Lambda (1405) \to \Lambda \gamma$ is suitable for the study of the
  $\bar{K} N$ component in $\Lambda (1405)$.  Fixing the $\Lambda
  (1405)$-$\pi \Sigma$ coupling constant from the usual decay of
  $\Lambda (1405) \to \pi \Sigma$, we show a relation between the
  absolute value of the $\bar{K} N$ compositeness for $\Lambda (1405)$
  and the radiative decay width of $\Lambda (1405) \to \Lambda \gamma$
  and $\Sigma ^{0} \gamma$, and we find that large decay width to
  $\Lambda \gamma$ implies large $\bar{K}N$ compositeness for $\Lambda
  (1405)$. By using the ``experimental'' data on the radiative decay
  widths, which is based on an isobar model fitting of the $K^{-}p$
  atom data, we estimate the $\bar{K}N$ compositeness for $\Lambda
  (1405)$.  We also discuss the pole position dependence of our
  relation on the $\Lambda (1405)$ radiative decay width and the
  effects of the two-pole structure for $\Lambda (1405)$.
\end{abstract}

% \keywords{Suggested keywords}%Use showkeys class option if keyword
\maketitle

\section{Introduction}

Determination of the internal structure of hadrons is one of the most
important issues in the physics of strong interaction.  Especially
exotic hadrons, which have different configurations from $q\bar{q}$
for mesons or $qqq$ for baryons, are of interest because there is no
clear experimental evidence on the existence of exotic hadrons while
the fundamental theory of strong interaction, quantum chromodynamics
(QCD), does not forbid such exotic hadrons~\cite{Beringer:1900zz}.
Among others, $\Lambda (1405)$ is a ``classical'' example of the
exotic hadron candidates.  One of the remarkable properties for
$\Lambda (1405)$ is its anomalously light mass.  Actually in the
$1/2^{-}$ state the lowest $\Lambda$ excited state, $\Lambda (1405)$,
is lighter than that of the nucleon excitation, $N(1535)$, although
$\Lambda (1405)$ has a strange quark, which is heavier than up and
down quarks.

As an interpretation of the $\Lambda (1405)$ properties, it has been
considered that $\Lambda (1405)$ should be a $\bar{K} N$ quasi-bound
state rather than an $uds$ three-quark state~\cite{Dalitz:1960du,
  Dalitz:1967fp}.  This is reasonable, because $\Lambda (1405)$ exists
just below the $\bar{K} N$ threshold and the interaction between
$\bar{K} N$ is strongly attractive in the isospin $I=0$ channel.  The
idea that $\Lambda (1405)$ should be a $\bar{K} N$ quasi-bound state
is recently reconfirmed by the so-called chiral unitary
approach~\cite{Kaiser:1995eg, Oset:1997it, Oller:2000fj, Lutz:2001yb,
  Jido:2003cb}, which is based on the coupled-channels scattering
unitarity with the interaction between hadrons restricted by the
spontaneous chiral symmetry breaking in QCD, {\it i.e.} chiral
perturbation theory.  The chiral unitary approach reproduces the
low-energy $\bar{K} N$ dynamics fairly well and dynamically generates
$\Lambda (1405)$ without introducing explicit poles.  In the
experimental side, precise measurements have been recently performed
so as to reveal the structure of $\Lambda (1405)$.  For example, in
Refs.~\cite{Niiyama:2008rt, Moriya:2013eb} the $\Lambda (1405)$ line
shape has been measured in the photoproduction, which is closely
related to the underlying dynamics and the internal
structure of $\Lambda (1405)$~\cite{Nacher:1998mi}.  In addition,
there are discussions that the internal structure of $\Lambda (1405)$
can be extracted from, {\it e.g.} the coalescence of $\Lambda (1405)$
in heavy-ion collisions~\cite{Cho:2010db, Cho:2011ew} and the
exclusive production with high momentum $\pi ^{-}$
beam~\cite{Kawamura:2013iia}.

For the determination of the internal structures of exotic hadron
candidates, one needs to pin down quantities which can be an evidence
on the exotic structure.  The hadron yields in heavy-ion
collisions~\cite{Cho:2010db, Cho:2011ew} and the
constituent-counting rule in high-energy exclusive
production~\cite{Kawamura:2013iia} are the examples.  In addition to
them, compositeness has been recently discussed as a quantity to
determine the amount of two-body components in hadron
resonances~\cite{Hyodo:2011qc, Aceti:2012dd, Xiao:2012vv,
  Aceti:2013jg, Hyodo:2013nka}.  Here the compositeness is defined as
a fraction of the two-body components in a resonance as well as a
bound state and can be evaluated from the squared coupling constant of
the resonance to the two-body states and a kinematical factor.
Originally, whether a particle is elementary or composite was
intensively discussed in 1960's in terms of the field renormalization
constant $Z$~\cite{Salam:1962ap, Weinberg:1962hj, Ezawa:1963zz}, which
measures the bare state contribution rather than the composite state.
A striking result is that a deuteron is indeed a proton-neutron bound
state as shown in Ref.~\cite{Weinberg:1965zz}, in which a relation
between the field renormalization constant of a deuteron and the
scattering length and effective range in the proton-neutron scattering
is established in the small binding energy limit in a model
independent way.  Then attempts to investigate the structures of
hadrons from the field renormalization constant $Z$ have been made in,
{\it e.g.}  Refs.~\cite{Baru:2003qq, Hanhart:2007cm, Hanhart:2011jz}.
Recently the concept of the compositeness has been discussed in the
context of the chiral unitary approach to determine internal
structures of dynamically generated states~\cite{Hyodo:2011qc}.
Although the compositeness could be a complex value for resonances, it
would be an evidence of dominance of a two-body composite state or an
elementary state, as not $\pi \pi$ state for the
$\rho$ meson~\cite{Aceti:2012dd} and not $K \pi$ state for
the $K^{\ast}$ meson~\cite{Xiao:2012vv}, and also for the baryon
decuplet~\cite{Aceti:2013jg}.

\begin{figure}[!t]
  \centering
  \begin{tabular}{cc}
    \PsfigII{0.22}{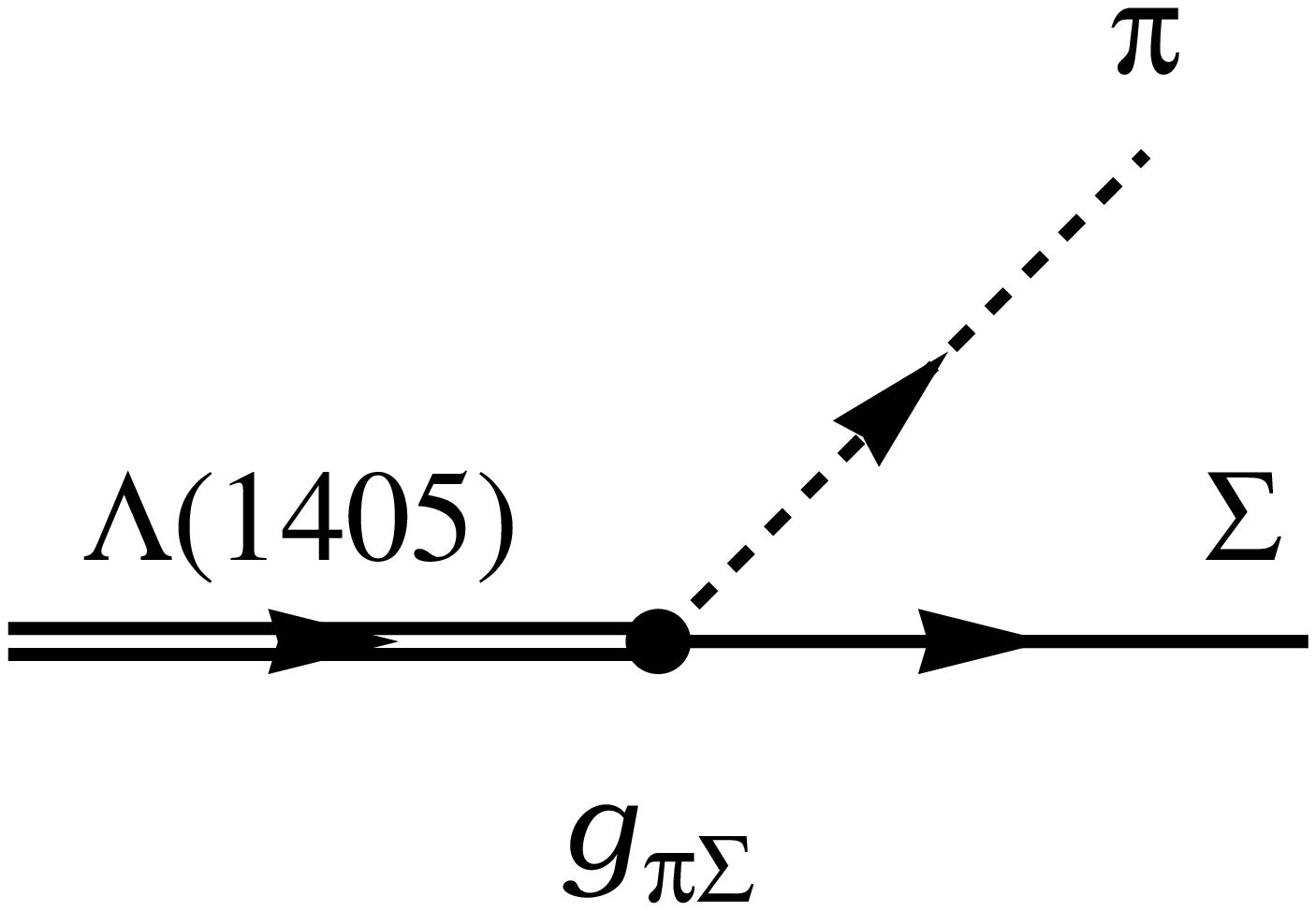} & 
    \PsfigII{0.22}{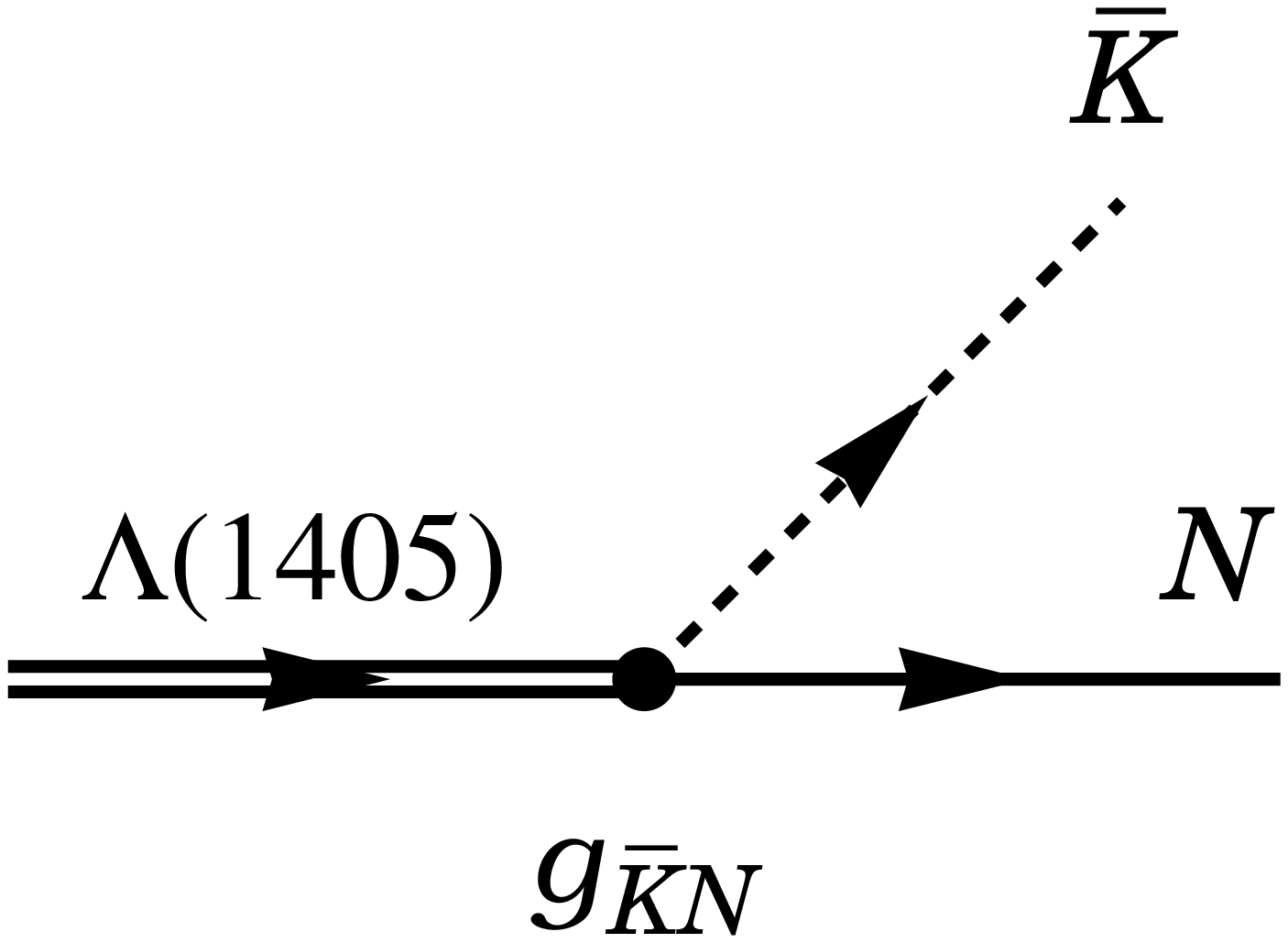} 
  \end{tabular}
  \caption{Couplings of $\Lambda (1405)$ to the $\pi \Sigma$ (left)
    and $\bar{K} N$ (right) states.  The $\Lambda (1405)$-$\pi \Sigma$
    coupling constant $g_{\pi \Sigma}$ can be extracted from the decay
    width of $\Lambda (1405) \to \pi \Sigma$, but one cannot extract
    the $\Lambda (1405)$-$\bar{K} N$ coupling constant $g_{\bar{K}N}$
    in a similar manner because $\Lambda (1405)$ exists below the
    $\bar{K} N$ threshold.}
 \label{fig:diag_coup}
\end{figure}

From these observations, we would like to determine the $\bar{K} N$
compositeness of $\Lambda (1405)$ from experimental information.
Since the compositeness of $\Lambda (1405)$ can be evaluated from the
coupling constant of $\Lambda (1405)$ (see Fig.~\ref{fig:diag_coup})
and a kinematic factor, {\it i.e.} derivation of the meson-baryon loop
integral on the $\Lambda (1405)$ pole position, in order to determine
the $\bar{K}N$ compositeness for $\Lambda (1405)$ we need to know the
coupling constant of $\Lambda (1405)$ to $\bar{K} N$ as well as its
pole position.  However, while we can easily extract the $\Lambda
(1405)$-$\pi \Sigma$ coupling constant from the $\Lambda (1405) \to
\pi \Sigma$ decay, the $\Lambda (1405)$-$\bar{K} N$ coupling constant
cannot be determined directly in experiments because $\Lambda (1405)$
exists below the $\bar{K} N$ threshold.  Therefore, in order to
determine the $\bar{K} N$ compositeness for $\Lambda (1405)$ we have
to consider reactions which are sensitive to the $\Lambda
(1405)$-$\bar{K} N$ coupling constant.

For such a reaction, we here treat the radiative decay of $\Lambda
(1405)$: $\Lambda (1405) \to \Lambda \gamma$ and $\Sigma ^{0}
\gamma$~\cite{Burkhardt:1991ms}.  Since the $\Lambda (1405)$ radiative
decay is closely related to the structure of $\Lambda (1405)$ as an E1
transition, up to now the radiative decay widths of $\Lambda (1405)$
have been evaluated in several models~\cite{Darewych:1983yw,
  Kaxiras:1985zv, Warns:1990xi, Umino:1992hi, Schat:1994gm,
  Bijker:2000gq, VanCauteren:2005sm, Yu:2006sc, Geng:2007hz,
  An:2010wb, Doring:2010rd}.  However, the decay widths have not been
interpreted from the viewpoint of the $\bar{K} N$ composite
contribution in detail.  Therefore, in this study we give a relation
between the $\bar{K}N$ compositeness for $\Lambda (1405)$ and the
$\Lambda (1405)$ radiative decay width.  Actually, as studied in
Ref.~\cite{Geng:2007hz}, the $\Lambda (1405)$ radiative decay takes
place mainly through the $\bar{K} N$ loop accompanied with the
$\Lambda (1405)$-$\bar{K} N$ coupling.  We will see that this fact is
important to establish a relation between the $\bar{K}N$ compositeness
for $\Lambda (1405)$ and the radiative decay width of $\Lambda (1405)
\to \Lambda \gamma$.

This paper is organized as follows.  In Sec.~\ref{sec:2} we develop
our formulation of the compositeness in the context of the chiral
unitary approach.  Then the radiative decay of the $\Lambda (1405)$
resonance is formulated in Sec.~\ref{sec:3}.  In Sec.~\ref{sec:4} we
investigate possibilities to observe compositeness both for a $\bar{K}
N$ bound state without strong decay and for the $\Lambda (1405)$
resonance.  Section~\ref{sec:5} is devoted to the conclusion of our
study.

\section{Chiral unitary approach and compositeness}
\label{sec:2}

Recently the concept of compositeness has been intensively discussed
in the context of the so-called chiral unitary
approach~\cite{Kaiser:1995eg, Oset:1997it, Oller:2000fj, Lutz:2001yb,
  Jido:2003cb} for dynamically generated states~\cite{Hyodo:2011qc,
  Aceti:2012dd, Xiao:2012vv, Aceti:2013jg, Hyodo:2013nka}.  Therefore,
we firstly review the formulation of the chiral unitary approach and
the relation to the compositeness.

In the chiral unitary approach we construct an $s$-wave
meson-baryon scattering amplitude $T_{ij}(s)$ by using the scattering
unitarity, which is expressed as
\begin{equation}
\text{Im} {T_{ij}}^{-1} ( s ) 
= \delta _{ij} \frac{\rho _{i} ( s )}{2}
\theta ( s - s_{i}^{\rm th} ) ,
\label{eq:unitarity}
\end{equation}
where $i$ and $j$ denote channels, $s$ is the Mandelstam variable,
$\theta (x)$ is the Heaviside step function, and $s_{i}^{\rm
  th}=(m_{i} + M_{i})^{2}$ is the threshold in $i$ channel with
$m_{i}$ and $M_{i}$ being meson and baryon masses in channel $i$,
respectively.  The $i$-channel meson-baryon phase space $\rho _{i}
(s)$ is defined as
\begin{equation}
\rho _{i} (s) \equiv 
\frac{M_{i} \lambda ^{1/2}(s, \, m_{i}^{2}, \, M_{i}^{2})}{4 \pi s} ,
\end{equation}
with the K\"{a}llen function $\lambda (x, \, y, \,
z)=x^{2}+y^{2}+z^{2}-2xy-2yz-2zx$. Then, with neglect of the left-hand
cut, one can construct a scattering equation in the Bethe-Salpeter
type from the expression of the scattering
unitarity~\eqref{eq:unitarity} by using the N/D
method~\cite{Oller:2000fj}:
\begin{equation}
T_{ij} (s) = V_{ij} (s) + \sum _{k} V_{ik} (s) G_{k} (s) T_{kj} (s) ,
\label{eq:BS}
\end{equation}
with an interaction kernel $V_{ij}$ which coincides with the
interactions taken from the chiral perturbation theory in the order
matching scheme, and the dispersion integral $G_{i}$:
\begin{equation}
G_{i} ( s ) 
= - \tilde{a}_{i}(s_{0}) - (s - s_{0})
\int _{s_{i}^{\rm th}}^{\infty} \frac{d s^{\prime}}{2 \pi}
\frac{\rho _{i} (s^{\prime})}{(s^{\prime} - s - i \epsilon) 
(s^{\prime} - s_{0})} .
\end{equation}
Here $\tilde{a}_{i}$ is a subtraction constant at certain energy
$s_{0}$, $\epsilon$ is an infinitesimal positive value, and $G_{i}(s)$
is known to equal, except for an infinite constant, the two-body loop
integral.  At the leading order of the chiral interaction for
$V_{ij}$, which corresponds to the Weinberg-Tomozawa interaction plus
$s$- and $u$-channel Born terms, only the subtraction constant in each
channel is the model parameter.  On the other hand, if one takes into
account the next-to-leading order for $V_{ij}$, the low-energy
constants in the interaction kernel also become model parameters.

By fitting the branching ratios of $K^{-}p$ at its threshold, the
chiral unitary approach can fairly well reproduce the existing
experimental cross sections of the low-energy $K^{-}p$ to various
meson-baryon channels even only with the Weinberg-Tomozawa
interaction~\cite{Kaiser:1995eg, Oset:1997it, Oller:2000fj}.
Furthermore, the chiral unitary approach dynamically generates the
$\Lambda (1405)$ resonance without introducing explicit resonance
poles~\cite{Jido:2003cb} and reproduces $\Lambda (1405)$ spectrum in
experiments.  This approach supports the meson-baryon bound state
picture for the $\Lambda (1405)$ resonance by revealing, {\it e.g.}  a
predominance of the meson-baryon component~\cite{Hyodo:2008xr}, its
large-$N_{c}$ scaling behavior~\cite{Hyodo:2007np, Roca:2008kr}, and
its spatial size~\cite{Sekihara:2008qk, Sekihara:2010uz,
  Sekihara:2012xp}.  Recent studies within the chiral unitary approach
as well as experimental conditions on $\Lambda (1405)$ are summarized
in the review article~\cite{Hyodo:2011ur}.

In addition, it is a prediction of the chiral unitary approach that
$\Lambda (1405)$ is a superposition of two resonance poles in the
energy region between $\pi \Sigma$ and $\bar{K} N$
thresholds~\cite{Jido:2003cb}.  One pole sitting in higher energy
around $1420 \mev$ shows dominant coupling to $\bar{K} N$ and is
expected to originate from the $\bar{K} N$ bound state while the other
lower pole with large imaginary part strongly couples to $\pi
\Sigma$~\cite{Hyodo:2007jq}.  The higher pole around $1420 \mev$ is of
interest because it means that one will observe $\Lambda (1405)$
spectrum which has a peak at $1420 \mev$ instead of the nominal $1405
\mev$ in the $\bar{K} N \to \pi \Sigma$ amplitude~\cite{Jido:2003cb}.
Indeed, an experiment of the $K^{-}d \to \pi \Sigma n$
reaction~\cite{Braun:1977wd} gives a support of the $\Lambda (1405)$
peak at $1420 \mev$ rather than $1405 \mev$, which has been confirmed
also by the theoretical calculations in the chiral unitary
approach~\cite{Jido:2009jf, Jido:2010rx, YamagataSekihara:2012yv,
  Jido:2012cy}.  Also we note that a very recent experiment of the
$\Lambda (1405)$ electroproduction~\cite{Lu:2013nza} gives the
$\Lambda (1405)$ line shape corresponding approximately to the
predictions of a two-pole picture for the $\Lambda (1405)$.

In general, the resonances as well as the bound states appear as poles
of the scattering amplitude in the complex $s$ plane.  Actually poles
of the dynamically generated states in the chiral unitary approach are
expressed as
\begin{equation}
T_{ij} ( s ) = \frac{g_{i} g_{j}}{\sqrt{s} - Z_{\rm pole}} 
+ T_{ij}^{\rm BG} . 
\label{eq:Tpole}
\end{equation}
where $g_{i}$ is the coupling constant of the dynamically generated
state to the channel $i$, $Z_{\rm pole}=M_{\rm R} - i \Gamma _{\rm
  R}/2$ is the pole position with $M_{\rm R}$ and $\Gamma _{\rm R}$
being interpreted as the mass and width of the state, respectively,
and $T_{ij}^{\rm BG}$ is a background term which is regular at
$\sqrt{s} \to Z_{\rm pole}$.  Here we have defined the coupling
constant $g_{i}$ as the residue of the resonance pole with respect to
$\sqrt{s}$ rather than $s$ so that the $\Lambda (1405)$ field has
dimension of mass to the three half in our notation.  Now, according
to the discussions on the scattering equations and field
renormalizations, it was implied in Ref.~\cite{Hyodo:2011qc} that the
compositeness of the dynamically generated resonances with respect to
the $i$ channel, $X_{i}$, is related to the coupling constant $g_{i}$
as
\begin{equation}
X_{i} = - g_{i}^{2} \frac{d G_{i}}{d \sqrt{s}} ( \sqrt{s} = Z_{\rm pole} ) . 
\label{eq:compositeness}
\end{equation}
The compositeness $X_{i}$ approaches unity if the system is dominated
by the $i$-channel two-body component, while it becomes zero if the
system does not contain the $i$-channel two-body component.  On the
other hand, we can define the elementarity $Z$, which corresponds to
the field renormalization constant and measures the fraction of the
bare state contribution rather than the two-body state, as the
residual part of the decomposition of unity:
\begin{equation}
Z = 1 - \sum _{i} X_{i} . 
\label{eq:unity}
\end{equation}
Then using the generalized Ward identity for the dynamically generated
states proved in Ref.~\cite{Sekihara:2010uz}
\begin{equation}
- \sum _{i, j} \left [ 
g_{i}^{2} \frac{d G_{i}}{d \sqrt{s}} \delta _{ij} 
+ g_{i} G_{i} \frac{d V_{ij}}{d \sqrt{s}} G_{j} g_{j} 
\right ]_{\sqrt{s} = Z_{\rm pole}} = 1 , 
\end{equation}
we can express the elementarity as
\begin{equation}
Z = - \sum _{i, j} \left . g_{i} G_{i} \frac{d V_{ij}}{d \sqrt{s}} G_{j} g_{j} 
\right | _{\rm \sqrt{s} = Z_{\rm pole}} .
\end{equation}
An important point to be noted is that the compositeness $X_{i}$ is
expressed as the squared coupling constant $g_{i}^{2}$, which contains
information on the dynamics, times the derivative of the dispersion
integral $G_{i}$, which depends only on the kinematics.  We also note
that each $X_{i}$ as well as the elementarity $Z$, which take real
values for bound states without decay width, become complex for
resonance states.  However, the sum of the compositeness and the
elementarity is exactly unity as in Eq.~\eqref{eq:unity}.  In this
sense we can extract the amount of the $i$ channel component for the
resonance states from the compositeness $X_{i}$.

At last of this section we demonstrate the $\pi \Sigma$ compositeness
of $\Lambda (1405)$ from the usual decay of $\Lambda (1405)$, $\Lambda
(1405) \to \pi \Sigma$, assuming the isospin symmetry both for the
coupling constant and the hadron masses.  The decay width $\Gamma
_{\Lambda (1405)}$ is related to the $\pi \Sigma$ coupling in the
particle basis, $g_{\pi \Sigma} =g_{\pi ^{+} \Sigma ^{-}} = g_{\pi
  ^{-} \Sigma ^{+}} = g_{\pi ^{0} \Sigma ^{0}}$, in the following form
\begin{equation}
\Gamma _{\Lambda (1405)} = 3 \times 
\frac{p_{\rm cm} M_{\Sigma}}{2 \pi M_{\Lambda (1405)}} | g_{\pi \Sigma} |^{2} , 
\label{eq:Gamma_piSigma}
\end{equation}
where $p_{\rm cm}$ is the center-of-mass three-momentum of the
final-state $\pi$, $M_{\Sigma}$ is the averaged $\Sigma$ baryon mass,
and the factor three corresponds to the three possible final states:
$\pi ^{+} \Sigma ^{-}$, $\pi ^{-} \Sigma ^{+}$, and $\pi ^{0} \Sigma
^{0}$.  From the experimental data $M_{\Lambda (1405)}=1405 \mev$ and
$\Gamma _{\Lambda (1405)}=50 \mev$ in Particle Data
Group~\cite{Beringer:1900zz}, we have $| g_{\pi \Sigma}| =0.91$.
Then, the absolute value of the $\pi \Sigma$ compositeness can be
evaluated as
\begin{equation}
| X _{\pi \Sigma} | = 3 \times \left | g_{\pi \Sigma}^{2}
\frac{d G_{\pi \Sigma}}{d \sqrt{s}} ( \sqrt{s} = Z_{\rm pole} 
) \right | 
= 0.19 , 
\end{equation}
where $Z_{\rm pole} = M_{\Lambda (1405)} - i \Gamma _{\Lambda (1405)}
/2$.\footnote{We note that the pole position of $\Lambda (1405)$,
  which is necessary to evaluate the
  compositeness~\eqref{eq:compositeness}, is not well determined from
  experiments.  Actually the interference between $\Lambda (1405)$ and
  the $I=1$ background could shift the $\Lambda (1405)$ peak position
  in the spectrum as in Refs.~\cite{Niiyama:2008rt, Moriya:2013eb,
    Nacher:1998mi} and moreover the chiral unitary approach predicts
  the two-pole structure for $\Lambda (1405)$~\cite{Jido:2003cb}.
  Here we simply assume that the pole position is obtained from the
  mass and width in the Particle Data Group.  In this study we will
  discuss the pole position dependence of the $\Lambda (1405)$
  radiative decay width and the effects of the two-pole structure for
  $\Lambda (1405)$.}  This means that the $\pi \Sigma$ component in
the $\Lambda (1405)$ resonance is not dominant and hence the $\Lambda
(1405)$ should originate some dynamics rather than the $\pi \Sigma$
interaction.  In a similar manner we could evaluate the $\bar{K} N$
compositeness of $\Lambda (1405)$, but it is impossible because
$\Lambda (1405)$ exists below the $\bar{K} N$ threshold and the decay
to $\bar{K} N$ is invisible.  Furthermore, there are no direct
relations between the $\bar{K}N$ compositeness and observables such as
the $K^{-}p$ scattering length in contrast to the deuteron case.
Therefore, in order to obtain information on the $\bar{K} N$ component
inside $\Lambda (1405)$, we have to observe other decay modes which
are sensitive to the $\Lambda (1405)$-$\bar{K} N$ coupling constant,
such as the radiative decay.

\section{Formulation of  radiative decay}
\label{sec:3}

\begin{figure*}[!ht]
  \centering
  \begin{tabular*}{\textwidth}{@{\extracolsep{\fill}}ccc}
    \PsfigII{0.22}{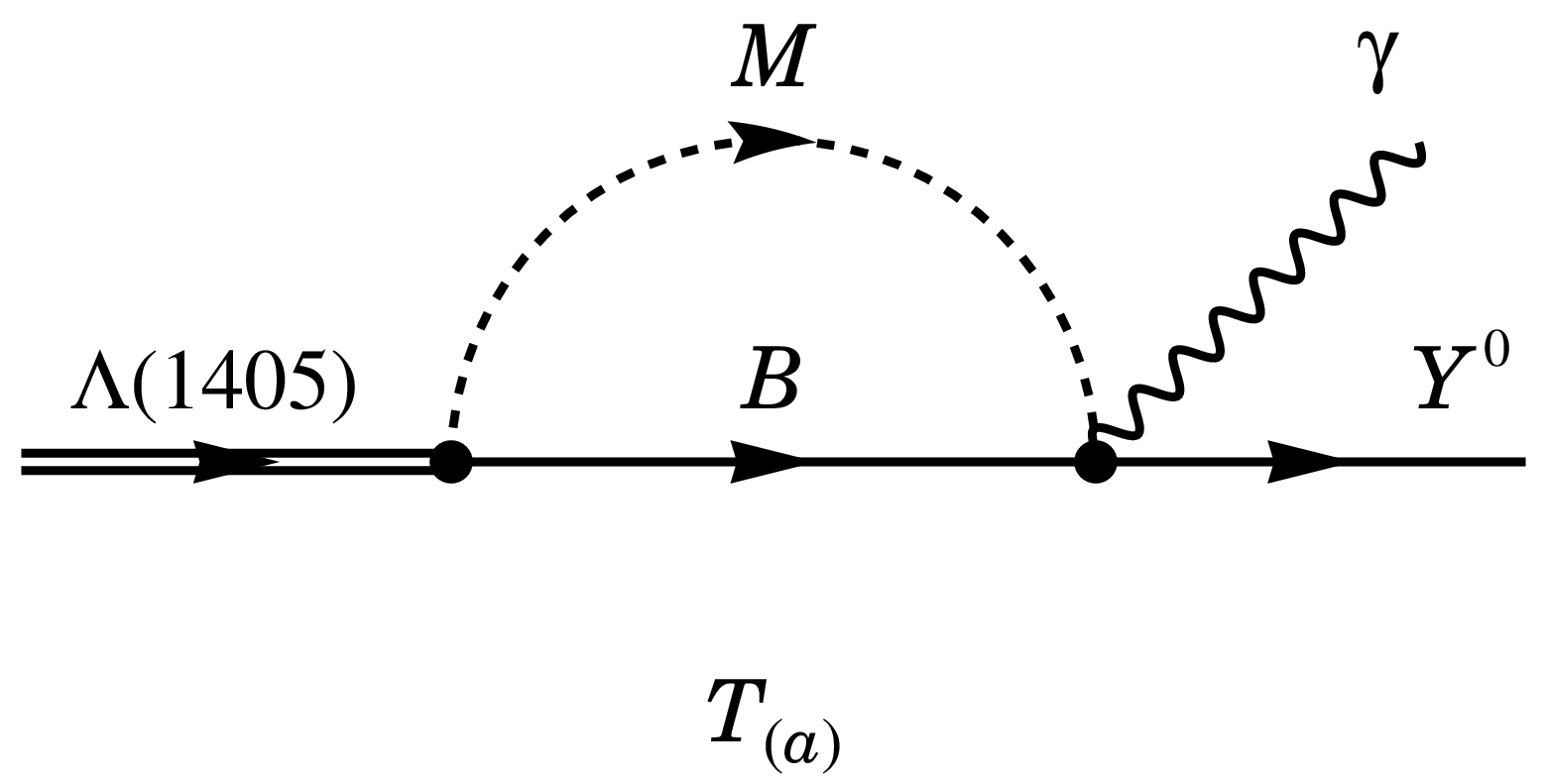} & 
    \PsfigII{0.22}{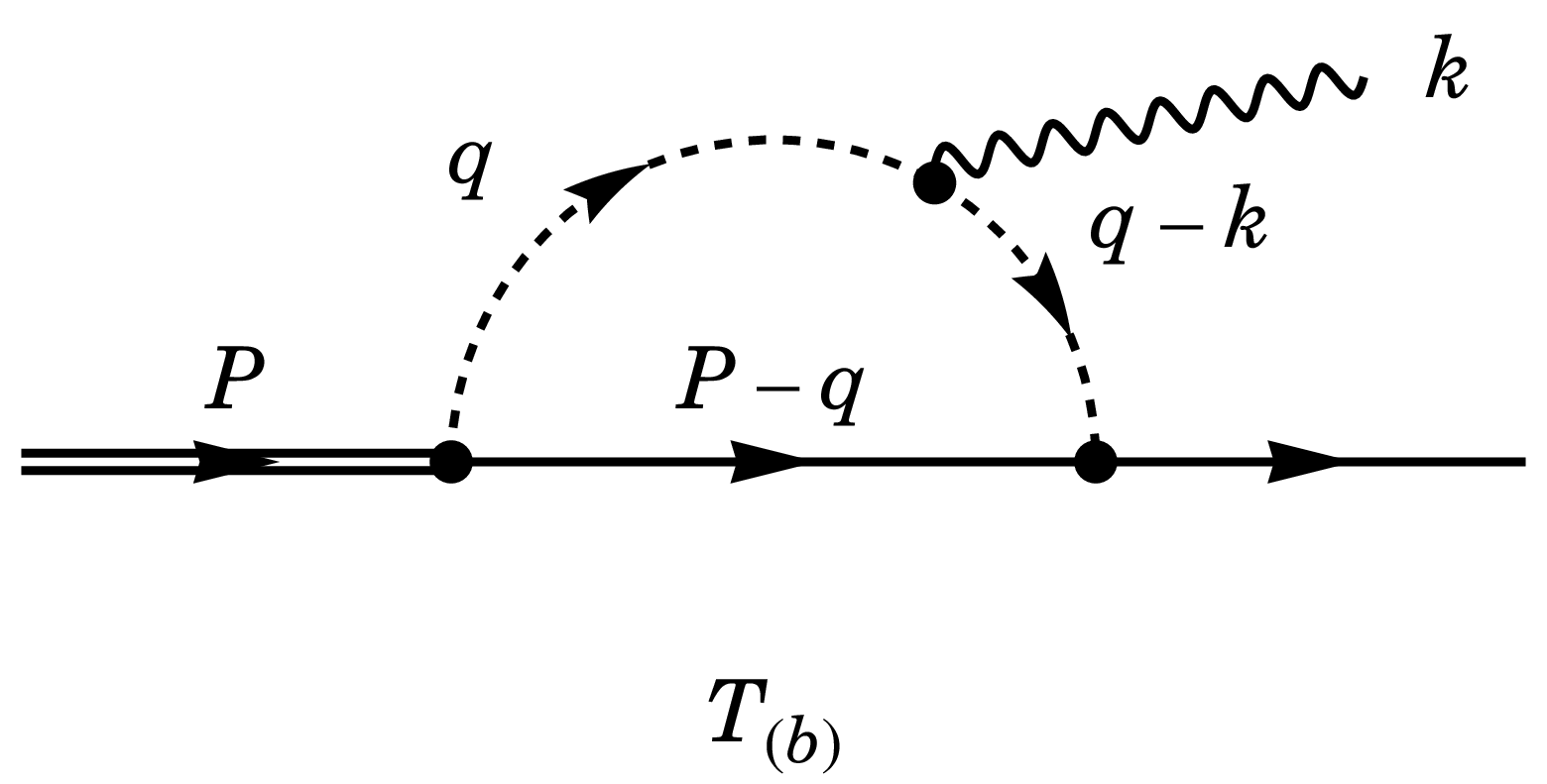} & 
    \PsfigII{0.22}{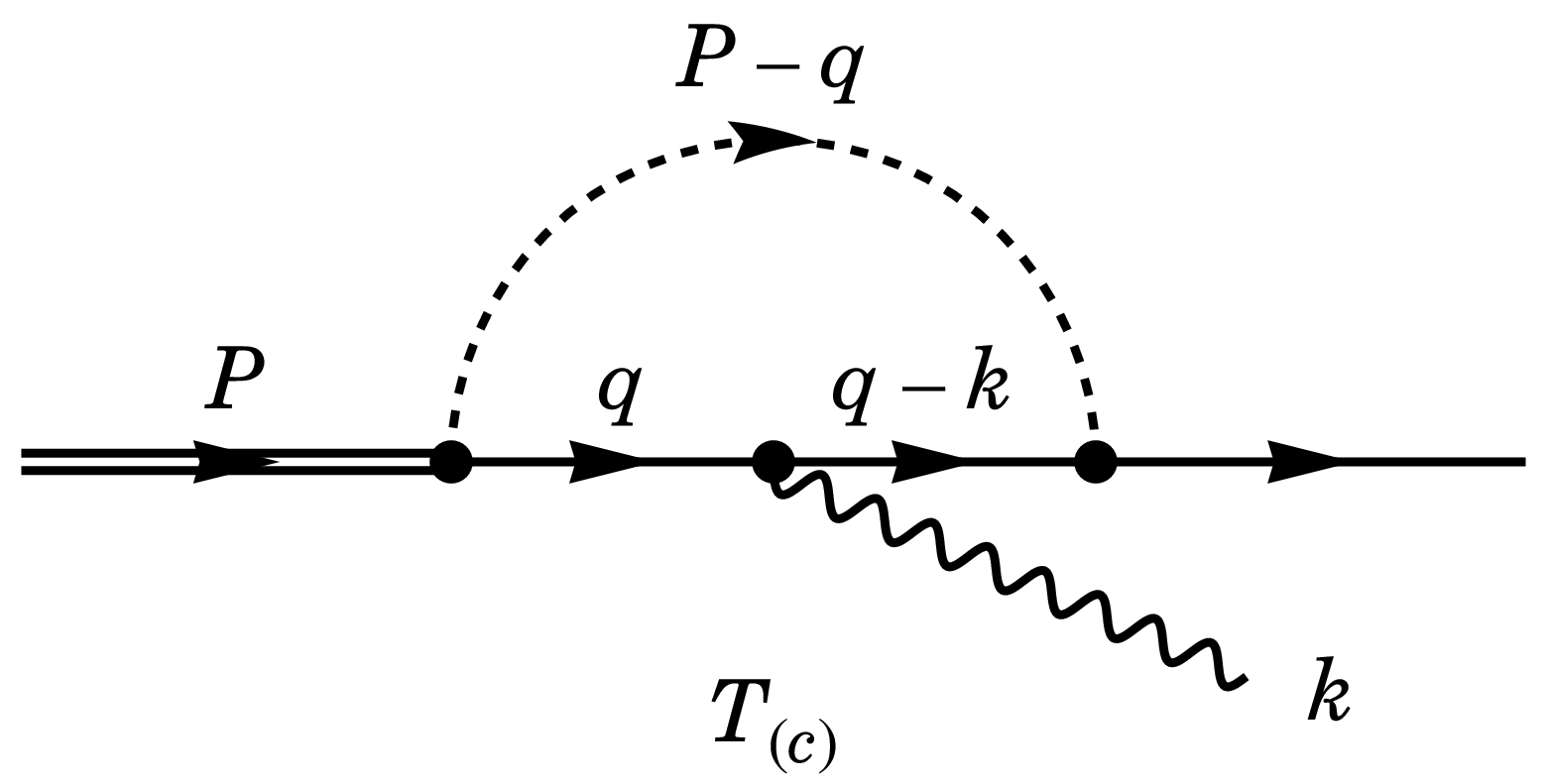} 
  \end{tabular*}
  \caption{Feynman diagrams for the $\Lambda (1405)$ radiative decay.
    Here $M$ and $B$ in the left diagram denote mesons and baryons,
    respectively, and $P$, $q$, $P-q$, $k$, and $q-k$ in the middle
    and right diagrams indicate the momenta carried by the
    corresponding mesons, baryons, and photons.}
  \label{fig:diag_rad}
\end{figure*}

Now let us formulate the radiative decay of $\Lambda (1405)$: $\Lambda
(1405) \to Y^{0} \gamma$ with $Y^{0}=\Lambda$, $\Sigma ^{0}$.  Our
formulation is based on that developed in Ref.~\cite{Geng:2007hz}.
Here the radiative decay widths are perturbatively calculated and
hence we use the same coupling constants in the formulation as those
obtained in the strong interaction without the electromagnetic
interaction, {\it i.e.} those in Eq.~\eqref{eq:Tpole}.  The relevant
diagrams for the radiative decay are shown in Fig.~\ref{fig:diag_rad},
and from the Feynman rules summarized in Appendix~\ref{sec:A}, each
diagram gives the decay amplitude as
\begin{equation}
- i T_{(a)} = - e \sigma _{\mu} \epsilon _{\nu}^{\ast} 
\sum _{i} g_{i} Q_{M_{i}} \tilde{V}_{i Y^{0}} 
g^{\mu \nu} G_{i} (P) , 
\label{eq:Ta}
\end{equation}
\begin{equation}
- i T_{(b)} = + e \sigma _{\mu} \epsilon _{\nu}^{\ast} 
\sum _{i} g_{i} Q_{M_{i}} \tilde{V}_{i Y^{0}} D_{i Y^{0} (1)}^{\mu \nu} ( P , \, k ) ,
\label{eq:Tb}
\end{equation}
\begin{equation}
- i T_{(c)} = + e \sigma _{\mu} \epsilon _{\nu}^{\ast} 
\sum _{i} g_{i} Q_{B_{i}} \tilde{V}_{i Y^{0}} D_{i Y^{0} (2)}^{\mu \nu} ( P , \, k ) ,
\label{eq:Tc}
\end{equation}
where $e$ is the elementary charge, $\sigma ^{\mu}$ is defined as
$\sigma ^{\mu} = (0, \, \bm{\sigma})$ with the Pauli matrices $\sigma
^{i}$ ($i=1, \, 2, \, 3$) for baryon spinors, and $\epsilon ^{\ast
  \mu}$ is the polarization of the final-state photon.  Inside the
summations with respect to the channel $i$, $g_{i}$ is the coupling
constant of $\Lambda (1405)$ to the channel $i$, $Q_{M_{i}}$ and
$Q_{B_{i}}$ are charges of the meson and baryon in channel $i$,
respectively, and $\tilde{V}_{i Y^{0}}$ is the meson-baryon-baryon
($MBB$) coupling strength:
\begin{equation}
\tilde{V}_{i Y^{0}} = \alpha _{i Y^{0}} \frac{D + F}{2 f} 
+ \beta _{i Y^{0}} \frac{D - F}{2 f} ,
\end{equation}
with $\SUN{3}$ coefficient $\alpha$ and $\beta$, parameters $D$ and
$F$, and the meson decay constant $f$.  The loop integrals $G_{i}$,
$D_{i Y^{0}(1)}^{\mu \nu}$, and $D_{i Y^{0}(2)}^{\mu \nu}$ are defined
as
\begin{equation}
G_{i} ( P ) \equiv i \int \frac{d ^{4} q}{(2 \pi)^{4}} 
\frac{1}{q^{2} - m_{i}^{2}} 
\frac{2 M_{i}}{(P - q)^{2} - M_{i}^{2}} , 
\label{eq:def_G}
\end{equation}
\begin{align}
& D_{i Y^{0}(1)}^{\mu \nu} ( P , \, k )
\nonumber \\ & \equiv i \int \frac{d ^{4} q}{(2 \pi)^{4}} 
\frac{( q - k )^{\mu} ( 2 q - k )^{\nu}}
{[(q - k)^{2} - m_{i}^{2}] (q^{2} - m_{i}^{2})} 
\frac{2 M_{i}}{(P - q)^{2} - M_{i}^{2}} , 
\label{eq:def_D1}
\end{align}
\begin{align}
& D_{i Y^{0}(2)}^{\mu \nu} ( P , \, k )
\nonumber \\ & \equiv i \int \frac{d ^{4} q}{(2 \pi)^{4}} 
\frac{2 M_{i} ( P - q )^{\mu} ( 2 q - k )^{\nu}}
{[(q - k)^{2} - M_{i}^{2}] (q^{2} - M_{i}^{2})} 
\frac{1}{(P - q)^{2} - m_{i}^{2}} , 
\label{eq:def_D2}
\end{align}
with $P^{\mu}$ and $k^{\mu}$ being the momenta of the initial-state
$\Lambda (1405)$ and final-state photon, respectively.  Here the
squared masses in the denominators are implicitly assumed to be added
by $-i \epsilon$, where $\epsilon$ is an infinitesimal positive value:
$m^{2}\to m^{2}-i\epsilon$ and $M^{2}\to M^{2}-i\epsilon$.  In this
study the relevant channels with nonzero charge are $K^{-}p$, $\pi
^{+} \Sigma ^{-}$, $\pi ^{-} \Sigma ^{+}$, and $K^{+} \Xi ^{-}$, and
the values of $\alpha _{i Y^{0}}$ and $\beta _{i Y^{0}}$ relevant to
this study are listed in Table~\ref{tab:1}.  Since we have a relation
$Q_{B_{i}} = - Q_{M_{i}}$ for the meson-baryon states coupled to the
zero-charge $\Lambda (1405)$, the decay amplitudes are collected to
give the total decay amplitude:
\begin{equation}
T_{\Lambda (1405) \to Y^{0} \gamma} = - i e \sigma _{\mu} \epsilon _{\nu}^{\ast} 
\sum_{i} g_{i} Q_{M_{i}} \tilde{V}_{i Y^{0}} H_{i Y^{0}}^{\mu \nu} (P , \, k) , 
\end{equation}
with the summation of the loop integrals: 
\begin{equation}
H_{i Y^{0}}^{\mu \nu} ( P , \, k ) 
\equiv g^{\mu \nu} G_{i} ( P ) 
- D_{i Y^{0} (1)}^{\mu \nu} ( P , \, k ) + D_{i Y^{0} (2)}^{\mu \nu} ( P , \, k ) . 
\end{equation}

\begin{table}
  \caption{$\SUN{3}$ coupling strengths for the $MBB$ vertex.}
  \label{tab:1}
  \begin{ruledtabular}
    \begin{tabular*}{8.6cm}{@{\extracolsep{\fill}}ccccc}
      $i$ & $K^{-} p$ 
      & $\pi ^{+} \Sigma ^{-}$ & $\pi ^{-} \Sigma ^{+}$ & $K^{+} \Xi ^{-}$ \\
      \hline
      $\alpha _{i \Lambda}$ & $- 2 / \sqrt{3}$ 
      & $1 / \sqrt{3}$ & $1 / \sqrt{3}$ & $1 / \sqrt{3}$ \\
      $\beta _{i \Lambda}$ & $1 / \sqrt{3}$ 
      & $1 / \sqrt{3}$ & $1 / \sqrt{3}$ & $-2 / \sqrt{3}$ \\
      $\alpha _{i \Sigma ^{0}}$ & $0$ & $1$ & $-1$ & $1$ \\
      $\beta _{i \Sigma ^{0}}$ & $1$ & $-1$ & $1$ & $0$ \\
    \end{tabular*}
  \end{ruledtabular}
\end{table}

For the evaluation of the loop integrals $H_{i}^{\mu \nu}$, we take
the strategy developed in Refs.~\cite{Geng:2007hz, Doring:2007rz,
  Close:1992ay, Oller:1998ia, Marco:1999df, Roca:2006am} based on the
symmetries.  Namely, since we have only $P^{\mu}$ and $k^{\mu}$ to
describe the Lorentz indices $\mu$ and $\nu$ with respect to
$H_{i}^{\mu \nu}$, the general form of $H_{i}^{\mu \nu}$ should be
expressed as
\begin{align}
H_{i Y^{0}}^{\mu \nu} 
= & a_{i Y^{0}} g^{\mu \nu} + b_{i Y^{0}} P^{\mu} P^{\nu} + c_{i Y^{0}} P^{\mu} k^{\nu} 
\nonumber \\ 
& + d_{i Y^{0}} k^{\mu} P^{\nu} + e_{i Y^{0}} k^{\mu} k^{\nu} . 
\end{align}
Then, since we consider an on-shell photon in the final state, we have
the transverse polarization: $\epsilon _{\nu}^{\ast} k^{\nu}=0$.  This
means that the terms proportional to $c_{i Y^{0}}$ and $e_{i Y^{0}}$
vanish and do not contribute to the radiative decay, hence we do not
consider them in the following.  Furthermore, since the Ward identity
constrains $k_{\nu} H_{i Y^{0}}^{\mu \nu}=0$ in each channel $i$ (see
Appendix~\ref{sec:B}), we have
\begin{equation}
a_{i Y^{0}} k^{\mu} + b_{i Y^{0}} P^{\mu} ( P \cdot k ) 
+ d_{i Y^{0}} k^{\mu} ( P \cdot k ) = 0 .
\end{equation}
Here we can take values of $P^{\mu}$ and $k^{\mu}$ independently, so
the equation implies $b_{i Y^{0}}=0$ and 
\begin{equation}
a_{i Y^{0}} + d_{i Y^{0}} ( P \cdot k ) = 0 .
\label{eq:ad_relation}
\end{equation}
This relation has a special meaning.  Suppose that we would like to
calculate the term $a_{i Y^{0}}$.  Then Eq.~\eqref{eq:ad_relation}
tells us that evaluating $a_{i Y^{0}}$ is equivalent to evaluating
$d_{i Y^{0}}$.  Furthermore, although each loop integral contributing
to $a_{i Y^{0}}$ such as $G_{i}(P)$ diverges,
Eq.~\eqref{eq:ad_relation} indicates that every divergence cancels
each other to be a finite value because $d_{i Y^{0}}$ is obviously
finite as one can see from the dimension counting.  At last we
consider $\Lambda (1405)$ at rest, $P^{\mu} = (M_{\Lambda (1405)},
\, \bm{0})$, and the Coulomb gauge $\epsilon ^{0}=0$, so the term
$d_{i Y^{0}}$ does not contribute to the total amplitude due to
$\epsilon _{\nu}^{\ast} P^{\nu}=0$ and hence we only need $a_{i
  Y^{0}}$.

From the above discussion, what we have to calculate is the term
proportional to $k^{\mu} P^{\nu}$ in $g^{\mu \nu} G_{i}(P)$, $D_{i
  Y^{0} (1)}^{\mu \nu}$, and $D_{i Y^{0} (2)}^{\mu \nu}$ to translate
$d_{i Y^{0}}$ into $a_{i Y^{0}}$ via Eq.~\eqref{eq:ad_relation}.
Since the first one does not have a factor of $k^{\mu} P^{\nu}$ but
$g^{\mu \nu}$, we may not evaluate the first one.  Thus we concentrate
on the second and third ones.  Using the Feynman parameterization
\begin{equation}
\frac{1}{A B C} = \int _{0}^{1} d x \int _{0}^{1} d y 
\frac{2 x}{[A x y + B x (1 - y) + C (1 - x)]^{3}} , 
\end{equation}
and the four dimensional integration
\begin{equation}
\int \frac{d^{4} q}{( 2 \pi )^{4}} \frac{i}{(q^{2} - S)^{3}}
= \frac{1}{32 \pi ^{2} S} , 
\end{equation}
we evaluate the term proportional to $k^{\mu} P^{\nu}$ in $D_{i
  Y^{0}(1)}^{\mu \nu}$ as
\begin{equation}
D_{i Y^{0} (1)}^{\mu \nu} |_{k^{\mu} P^{\nu}} = 
- \frac{M_{i} k^{\mu} P^{\nu}}{4 \pi ^{2}} 
\int _{0}^{1} d x \int _{0}^{1} d y 
\frac{x ( 1 - x ) ( 1 - xy )}{S ( x , \, y )} , 
\end{equation}
where $S(x, \, y)$ is defined as
\begin{equation}
S ( x , \, y ) \equiv 
x m^{2} + ( 1 - x ) M^{2} - x ( 1 - x ) P^{2} + 2 x ( 1 - x ) y P \cdot k .
\end{equation}
In a similar manner, the term proportional to $k^{\mu} P^{\nu}$ in
$D_{i Y^{0} (2)}^{\mu \nu}$ is evaluated as
\begin{equation}
D_{i Y^{0} (2)}^{\mu \nu} |_{k^{\mu} P^{\nu}} = 
- \frac{M_{i} k^{\mu} P^{\nu}}{4 \pi ^{2}} 
\int _{0}^{1} d x \int _{0}^{1} d y 
\frac{x ( 1 - x )^{2} y}{S ( x , \, y )} .
\end{equation}
As a result, we have
\begin{align}
& d_{i Y^{0}} ( \sqrt{s} ) =
\frac{M_{i}}{4 \pi ^{2}} 
\int _{0}^{1} d x \int _{0}^{1} d y 
\frac{x ( 1 - x ) ( 1 - y )}{S ( x , \, y )}
\nonumber \\
& = \frac{M_{i}}{4 \pi ^{2}} \frac{1}{2 P \cdot k} 
\int _{0}^{1} d x \left [ 
- 1 + (1 - y_{0}) \log \left ( \frac{1 - y_{0}}{- y_{0}} 
\right ) \right ] , 
\label{eq:term_d}
\end{align}
where $\sqrt{s}$ is the total energy, $P^{\mu}=(\sqrt{s}, \, \bm{0})$,
to be fixed as the $\Lambda (1405)$ mass $\sqrt{s}=M_{\Lambda (1405)}$
and the $y$ integration is performed in the last line of the equation
and $y_{0}$ is defined as
\begin{equation}
y_{0} (\sqrt{s}) \equiv 
- \frac{x m^{2} + ( 1 - x ) M^{2} - x ( 1 - x ) s}
{2 x ( 1 - x ) P \cdot k} .
\end{equation}
We note that the logarithmic term in Eq.~\eqref{eq:term_d} generates
an imaginary part for $0<y_{0}<1$.  Then, by using
Eq.~\eqref{eq:ad_relation} $a_{i Y^{0}}$ is evaluated as
\begin{equation}
a_{i Y^{0}} (\sqrt{s}) = - 
\frac{M_{i}}{8 \pi ^{2}} 
\int _{0}^{1} d x \left [ 
- 1 + (1 - y_{0}) \log \left ( \frac{1 - y_{0}}{- y_{0}} 
\right ) \right ] .
\label{eq:form_ai}
\end{equation}
As a consequence, the total decay amplitude at the $\Lambda (1405)$
rest frame in the Coulomb gauge becomes
\begin{equation}
T_{\Lambda (1405) \to Y^{0} \gamma} 
= i \bm{\sigma} \cdot \bm{\epsilon}^{\ast} W_{Y^{0} \gamma}  ( \sqrt{s} ) ,
\end{equation}
with
\begin{equation}
W_{Y^{0} \gamma} ( \sqrt{s} )
\equiv 
e \sum _{i} g_{i} Q_{M_{i}} \tilde{V}_{i Y^{0}} a_{i Y^{0}} (\sqrt{s}) .
\end{equation}

Finally the radiative decay width is expressed as
\begin{equation}
\Gamma _{Y^{0} \gamma} 
= \frac{p_{\rm cm}^{\prime} M_{Y^{0}}}{\pi M_{\Lambda (1405)}}
| W_{Y^{0} \gamma} ( \sqrt{s} = M_{\Lambda (1405)} ) |^{2} ,
\label{eq:Gamma_rad}
\end{equation}
where $p_{\rm cm}^{\prime}$ is the center-of-mass three-momentum of
the final-state photon.  We note again that each loop integral,
$G(P)$, $D_{i(1)}^{\mu \nu}$, or $D_{i(2)}^{\mu \nu}$, diverges,
but the sum of them gives a finite value as seen in
Eq.~\eqref{eq:form_ai}.  Throughout this study we take $D+F=1.26$,
$D-F=0.33$, and $f=1.15 f_{\pi}$ with the pion decay constant
$f_{\pi}=93 \mev$.

\section{Results}
\label{sec:4}

In this section we calculate the radiative decay width of $\Lambda
(1405)$ as a function of the $\bar{K}N$ compositeness $X_{\bar{K}N}$.
Here we use the physical hadron masses, which breaks slightly the
isospin symmetry, but assume the isospin symmetry for the coupling
constants of $\Lambda (1405)$ to each channel except for the coupling
constants evaluated in the chiral unitary approach.  In order to
understand the behavior of the decay width, we firstly consider a
$\bar{K}N$ bound state in the $I=0$ channel without couplings to other
channels in Sec.~\ref{sec:4a}, and then the radiative decay width of
$\Lambda (1405)$ is evaluated in Sec.~\ref{sec:4b}.  We also discuss
the pole position dependence of the $\Lambda (1405)$ radiative decay
width and the effects of the two-pole structure for $\Lambda (1405)$
in Sec.~\ref{sec:4c}.

\subsection{Radiative decay of a bound state}
\label{sec:4a}

We firstly consider a $\bar{K}N (I=0)$ bound state by taking into
account only the $K^{-}p$ and $\bar{K}^{0}n$ channels while switching
off the couplings to other channels such as $\pi \Sigma$.  Hence, the
state is described only by $K^{-}p$ and $\bar{K}^{0}n$, and the
radiative decay takes place only through the $K^{-}p$ loop.  Here we
assume the isospin symmetry for the coupling constant: $g_{\bar{K}N}=
g_{K^{-}p}=g_{\bar{K}^{0}n}$.  We take $M_{\rm B}$ as the mass of the
$\bar{K}N$ bound state, which is real and positive, and measure the
binding energy $B_{\rm E}$ from the mean value of the $K^{-}p$ and
$\bar{K}^{0}n$ thresholds:
\begin{equation}
B_{\rm E} = \frac{M_{K^{-}} + M_{p} + M_{\bar{K}^{0}} + M_{n}}{2} - M_{\rm B} .
\end{equation}
In this condition the $\bar{K}N$ compositeness for the bound
state, which takes a real value, can be evaluated as
\begin{equation}
X_{\bar{K}N} = - g_{\bar{K}N}^{2} 
\left [ \frac{d G_{K^{-} p}}{d \sqrt{s}} + \frac{d G_{\bar{K}^{0} n}}{d \sqrt{s}} 
\right ]_{\sqrt{s} = M_{\rm B}} . 
\label{eq:comp_KN-I}
\end{equation}
This indicates that for a given binding energy the coupling constant
$g_{\bar{K}N}$ and the compositeness $X_{\bar{K}N}$ have one-to-one
correspondence.  In addition, since the model parameter of the radiative
decay width is only the coupling constant $g_{\bar{K}N}$, we have
one-to-one correspondence between the $\bar{K}N$ compositeness and the
radiative decay width as well as the $\bar{K}N$ coupling constant and
the radiative decay width.  Therefore, one can evaluate the radiative
decay width as a function of the $\bar{K}N$ compositeness
$X_{\bar{K}N}$ through the relation~\eqref{eq:comp_KN-I}.  Since the
mass of $\Lambda (1405)$ is $1405 \mev$ from the Particle Data
Group~\cite{Beringer:1900zz} while the chiral unitary approach
suggests a real part of the $\Lambda (1405)$ pole position around $1420
\mev$~\cite{Jido:2003cb}, we choose two binding energies $B_{\rm E}=10
\mev$ ($M_{\rm B}=1425 \mev$) and $B_{\rm E}=30 \mev$ ($M_{\rm B}=1405
\mev$).

\begin{figure}[!t]
  \centering
  \Psfig{8.6cm}{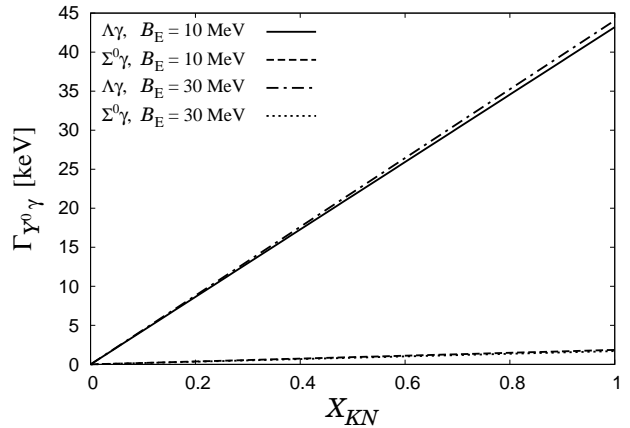}
  \caption{Radiative decay width of a $\bar{K} N$ bound state as a
    function of the $\bar{K}N$ compositeness.}
  \label{fig:rad_width_BS}
\end{figure}

In Fig.~\ref{fig:rad_width_BS} we show the radiative decay width of
the $\bar{K}N(I=0)$ bound state as a function of the $\bar{K}N$
compositeness with two binding energies $B_{\rm E}=10$ and $30 \mev$.
As on can see, the radiative decay width both to $\Lambda \gamma$ and
$\Sigma ^{0} \gamma$ is proportional to the $\bar{K}N$ compositeness
$X_{\bar{K}N}$ since both the decay width and the compositeness are
proportional to the squared coupling constant $g_{\bar{K}N}^{2}$.
Furthermore, it is interesting that the decay to $\Lambda \gamma$ is
dominant while the decay to $\Sigma ^{0} \gamma$ is quite small.  This
behavior can be understood by the coupling strengths of
$K^{-}p\Lambda$ and $K^{-}p\Sigma ^{0}$.  Namely, the flavor $\SUN{3}$
symmetry gives the coupling strengths:
\begin{equation}
\tilde{V}_{K^{-}p\Lambda} 
= - \frac{D + 3F}{2 \sqrt{3} f} 
\approx - \frac{0.63}{f}, 
\label{eq:VKpL}
\end{equation}
\begin{equation}
\tilde{V}_{K^{-}p\Sigma ^{0}} = \frac{D - F}{2 f} 
\approx \frac{0.17}{f}, 
\label{eq:VKpS}
\end{equation}
This difference gives a very large ratio $\sim 14$ for the radiative
decay width $\Gamma _{\Lambda \gamma} / \Gamma _{\Sigma ^{0} \gamma}$.
Figure~\ref{fig:rad_width_BS} means that we have established a
relation between the $\bar{K}N$ compositeness inside the $\bar{K}N
(I=0)$ bound state and the radiative decay width.

On the other hand, the binding energy dependence of the radiative
decay width is very small and almost invisible.  This indicates that
the behavior of the squared loop integral $a_{K^{-}pY^{0}}^{2}$ with
respect to the energy $\sqrt{s}$ is very similar to that of
$-dG_{K^{-}p}/d\sqrt{s}$ and $-dG_{\bar{K}^{0}n}/d\sqrt{s}$ in the
energy region $1400 \mev < \sqrt{s} < 1430 \mev$, and hence for a
given $X_{\bar{K}N}$ [see Eq.~\eqref{eq:comp_KN-I}] at
$\sqrt{s}=M_{\rm B}$ the radiative decay width $\Gamma \propto
g_{\bar{K}N}^{2} a_{K^{-}pY^{0}}^{2}(\sqrt{s}=M_{\rm B})$ takes almost
similar values independently of the bound state mass $M_{\rm B}$.  In
other words, since both $G$ and $dG/d\sqrt{s} (<0)$ monotonically
decrease as functions of $\sqrt{s}$ below the threshold, for a fixed
$X_{\bar{K}N}$ a larger binding energy leads to smaller $-dG/d\sqrt{s}$
and hence larger $g_{\bar{K}N}^{2}$.  Then, since the energy
dependence of $a_{K^{-}pY^{0}}^{2}$ is very similar to that of
$-dG_{K^{-}p}/d\sqrt{s}-dG_{\bar{K}^{0}n}/d\sqrt{s}$, a larger binding
energy leads to the smaller $a_{K^{-}pY^{0}}^{2}$ and hence for a
fixed $X_{\bar{K}N}$ the binding energy dependence of
$g_{\bar{K}N}^{2}a_{K^{-}pY^{0}}^{2}$ is almost cancelled.  This fact
will be more important when we consider the $\Lambda (1405)$
resonance.  Namely, we expect that the radiative decay width of
$\Lambda (1405)$ can be evaluated as a function of the $\bar{K} N$
compositeness almost independently of the $\Lambda (1405)$ pole
position, which is not well determined experimentally at present.

\subsection{Radiative decay of $\bm{\Lambda (1405)}$}
\label{sec:4b}

In the previous subsection we have studied the radiative decay of a
$\bar{K}N(I=0)$ bound state by taking into account only the $K^{-}p$
and $\bar{K}^{0}n$ channels while switching off the couplings to other
channels.  As a result we have established a relation between the
$\bar{K}N$ compositeness for the $\bar{K}N$ bound state and the
radiative decay width.  In this subsection we extend our discussion to
the $\Lambda (1405)$ resonance in multi-channels and investigate
whether or not we can establish a relation between the $\bar{K}N$
compositeness for $\Lambda (1405)$ and the $\Lambda (1405)$ radiative
decay width.

\begin{table}
  \caption{Radiative decay width of $\Lambda (1405)$ in the chiral 
    unitary approach with the recently updated 
    parameters~\cite{Ikeda:2011pi, 
      Ikeda:2012au}.}
  \label{tab:2}
  \begin{ruledtabular}
    \begin{tabular*}{8.6cm}{@{\extracolsep{\fill}}lcc}
      & $\Lambda (1405)$, higher pole & $\Lambda (1405)$, lower pole \\
      \hline
      $Z_{\rm pole}$ [MeV] & $1424 - 26 i$ & $1381 - 81i$ \\
      $g_{K^{-}p}$ & $2.25 + 0.87 i$ & $0.91 - 1.89 i$ \\
      $g_{\pi ^{+}\Sigma^{-}}$ & $0.57 + 1.00 i$ & $1.37 - 1.28 i$ \\
      $g_{\pi ^{-}\Sigma^{+}}$ & $0.62 + 1.06 i$ & $1.48 - 1.28 i$ \\
      $g_{K^{+}\Xi^{-}}$ & $0.23 + 0.08 i$ & $0.02 - 0.26 i$ \\
      $\Gamma _{\Lambda \gamma}$ [keV] & $96$ & $31$ \\
      $\Gamma _{\Sigma ^{0} \gamma}$ [keV] & $60$ & $94$ \\
    \end{tabular*}
  \end{ruledtabular}
\end{table}

Before evaluating the radiative decay width as a function of the
$\bar{K}N$ compositeness for $\Lambda (1405)$, we firstly evaluate the
$\Lambda (1405)$ radiative decay width in the chiral unitary approach
with the recently updated parameters~\cite{Ikeda:2011pi, Ikeda:2012au}.
In Refs.~\cite{Ikeda:2011pi, Ikeda:2012au} two poles corresponding to
$\Lambda (1405)$ are reconfirmed, and the pole positions of $\Lambda
(1405)$, the coupling constants to meson-baryon channels on the pole
positions, and the resulting radiative decay widths, which correspond
to the updated values with respect to the previous
study~\cite{Geng:2007hz}, are listed in Table~\ref{tab:2}.  As one can
see, the radiative decay width to $\Lambda \gamma$ is larger than that
to $\Sigma ^{0} \gamma$ for the higher $\Lambda (1405)$ pole whereas
decay to $\Sigma ^{0} \gamma$ is dominant for the lower pole.  This
behavior comes from the structure of the meson-baryon-baryon coupling
$\tilde{V}$ as discussed in Ref.~\cite{Geng:2007hz}.  Namely, the
$K^{-}p\Lambda$ coupling strength $\tilde{V}_{K^{-}p\Lambda}$ is large
compared to the $K^{-}p\Sigma ^{0}$ one $\tilde{V}_{K^{-}p\Sigma
  ^{0}}$, as we have already shown in Eqs.~\eqref{eq:VKpL} and
\eqref{eq:VKpS} in the previous subsection.  On the other hand, the
$\pi ^{\pm} \Sigma ^{\mp} \Lambda$ couplings are found to be
\begin{equation}
\tilde{V}_{\pi ^{+} \Sigma ^{-} \Lambda} 
= \tilde{V}_{\pi ^{-} \Sigma ^{+} \Lambda} 
= \frac{D}{\sqrt{3} f}
\approx \frac{0.46}{f} , 
\end{equation}
and hence, due to the opposite sign of the charge $Q_{\pi
  ^{+}}=-Q_{\pi ^{-}}=1$, for the $\pi \Sigma$ component tiny isospin
breakings in the loop integral $a_{\pi ^{\pm} \Sigma ^{\mp}}$ and in
the coupling constant $g_{\pi \Sigma}$ can contribute to the decay to
$\Lambda \gamma$.  Therefore, the decay to $\Lambda \gamma$ is
dominated by the $\bar{K}N$ component.  Similarly, for the decay to
$\Sigma ^{0} \gamma$ we have
\begin{equation}
\tilde{V}_{\pi ^{+} \Sigma ^{-} \Sigma ^{0}}
= - \tilde{V}_{\pi ^{-} \Sigma ^{+} \Sigma ^{0}}
= \frac{F}{f} 
\approx \frac{0.47}{f} , 
\end{equation}
which are larger than the $K^{-}p\Sigma ^{0}$ coupling
strength~\eqref{eq:VKpS}, and the constructive interference between
$\pi ^{\pm} \Sigma ^{\mp}$ does take place.  As a result, the $\Sigma
^{0} \gamma$ decay is dominated by the $\pi \Sigma$ component.  Then,
as listed in Table~\ref{tab:2}, the higher pole dominantly couples to
the $\bar{K}N$ channel while the lower pole strongly couples to the
$\pi \Sigma$ channel.  These coupling strengths lead to the large
$\Gamma _{\Lambda \gamma} / \Gamma _{\Sigma ^{0} \gamma}$ in the
higher pole and to the small $\Gamma _{\Lambda \gamma} / \Gamma
_{\Sigma ^{0} \gamma}$ in the lower pole.  We note that the
interference between $K^{-}p$ and $\pi ^{\pm} \Sigma ^{\mp}$ in
$\Gamma _{\Lambda \gamma}$ is constructive both for the higher and
lower $\Lambda (1405)$ poles, while the interference in $\Gamma
_{\Sigma ^{0} \gamma}$ is constructive (destructive) for the higher
(lower) pole.  In addition, we also note that for both poles the
couplings to the $K\Xi$ channel are small and hence the $K\Xi$
component can only scarcely contribute to the radiative decay.

Bearing these discussions in mind, we calculate the radiative decay
width of $\Lambda (1405)$ as a function of the $\bar{K}N$
compositeness for $\Lambda (1405)$.  Here we firstly fix the $\Lambda
(1405)$ pole position by using the mass and width taken from the
Particle Data Group~\cite{Beringer:1900zz}: $Z_{\rm pole}=M_{\Lambda
  (1405)} - i \Gamma _{\Lambda (1405)}/2$ with $M_{\Lambda
  (1405)}=1405 \mev$ and $\Gamma _{\Lambda (1405)} = 50 \mev$.  Then
we will discuss later the pole position dependence of the relation
between the radiative decay width and the $\bar{K}N$ compositeness.
The $\bar{K}N$ compositeness $X_{\bar{K}N}$ is related to the
$\bar{K}N$ coupling constant in the particle basis, $g_{\bar{K}N}
= g_{K^{-}p} = g_{\bar{K}^{0}n}$, as:
\begin{equation}
X_{\bar{K}N} = - g_{\bar{K}N}^{2} 
\left [ \frac{d G_{K^{-} p}}{d \sqrt{s}} + \frac{d G_{\bar{K}^{0} n}}{d \sqrt{s}} 
\right ]_{\sqrt{s} = Z_{\rm pole}} . 
\label{eq:comp_KN-II}
\end{equation}
However, since $\Lambda (1405)$ is a resonance state, the
compositeness $X_{\bar{K}N}$ as well as the coupling constant
$g_{\bar{K}N}$ are in general complex.  Therefore, in this study, in
order to evaluate the radiative decay width as a function of a real
variable, we use Eq.~\eqref{eq:comp_KN-II} to relate the absolute
value of the compositeness $|X_{\bar{K}N}|$ and that of the coupling
constant $|g_{\bar{K}N}|$.  We note that, although the absolute value
of the compositeness $|X_{\bar{K}N}|$ as well as the complex
compositeness for the $\Lambda (1405)$ resonance cannot be interpreted
as a probability of finding the $\bar{K}N$ component, it will be a
helpful information and be a guide to elucidating the structure of
$\Lambda (1405)$.  For instance, the large absolute value of the
$\bar{K}N$ compositeness $|X_{\bar{K}N}|$ is a necessary condition for
the $\bar{K}N$ bound state picture for $\Lambda (1405)$.  In this
strategy, for a given $|X_{\bar{K}N}|$ we can uniquely determine
$|g_{\bar{K}N}|$.  On the other hand, the $\pi \Sigma$ coupling
constant can be fixed from the usual $\Lambda (1405) \to \pi \Sigma$
decay, and from Eq.~\eqref{eq:Gamma_piSigma} we take $|g_{\pi
  \Sigma}|=0.91$.  Finally, since the $K\Xi$ component inside $\Lambda
(1405)$ should be small, we neglect the $K\Xi$ coupling: $g_{K^{+}\Xi
  ^{-}}=g_{K^{0}\Xi ^{0}}=0$.

Although we have fixed the absolute values of the coupling constants
$|g_{\bar{K}N}|$ and $|g_{\pi \Sigma}|$, the relative phase between
$g_{\bar{K}N}$ and $g_{\pi \Sigma}$ is not known and hence one cannot
calculate the interference term.\footnote{Since we assume the isospin
  symmetry for the coupling constants, $g_{\pi \Sigma} = g_{\pi ^{+}
    \Sigma ^{-}} = g_{\pi ^{-} \Sigma ^{+}} = g_{\pi ^{0} \Sigma
    ^{0}}$ and so on, the relative phase between $\pi ^{+} \Sigma
  ^{-}$ and $\pi ^{-} \Sigma ^{+}$ is already determined.}  Therefore,
in order to evaluate the radiative decay width, we take a procedure to
calculate both the maximally constructive and maximally destructive
terms.  Namely, we calculate decay amplitudes of:
\begin{align}
& W_{Y^{0}\gamma}^{\pm} = 
e \Big ( 
| g_{\bar{K}N} | \times 
\left | \tilde{V}_{K^{-} p Y^{0}} a_{K^{-} p Y^{0}} \right | 
\nonumber \\ & 
\pm | g_{\pi \Sigma} | \times 
\left | \tilde{V}_{\pi ^{+} \Sigma ^{-} Y^{0}} a_{\pi ^{+} \Sigma ^{-}  Y^{0}} 
- \tilde{V}_{\pi ^{-} \Sigma ^{+} Y^{0}} a_{\pi ^{-} \Sigma ^{+}  Y^{0}} \right | 
\Big ) 
\end{align}
and evaluate the decay width~\eqref{eq:Gamma_rad} so as to show
allowed range for the radiative decay width for each absolute value of
the $\bar{K}N$ compositeness $|X_{\bar{K}N}|$.

\begin{figure}[!t]
  \centering
  \Psfig{8.6cm}{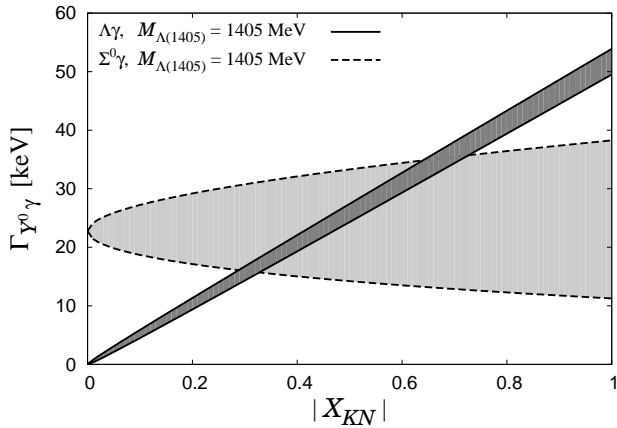}
  \caption{Radiative decay width of $\Lambda (1405)$ as a function of
    the absolute value of the $\bar{K}N$ compositeness. The $\Lambda
    (1405)$ mass is fixed as $M_{\Lambda (1405)}=1405 \mev$.}
  \label{fig:rad_width_Lam}
\end{figure}

The results of the allowed range of the $\Lambda (1405)$ radiative
decay widths are shown in Fig.~\ref{fig:rad_width_Lam} as functions of
the absolute value of the $\bar{K}N$ compositeness $|X_{\bar{K}N}|$.
As one can see from the figure, the range of the radiative decay width
to $\Lambda \gamma$ increases almost linearly with a small band as
$|X_{\bar{K}N}|$ increases.  This is because the $\pi ^{+} \Sigma
^{-}$ and $\pi ^{-} \Sigma ^{+}$ components largely cancel each other
and only a tiny isospin breaking part can contribute to the $\Lambda
\gamma$ decay.  This fact indicates that the radiative decay $\Lambda
(1405) \to \Lambda \gamma$ is suited to study the $\bar{K}N$
component inside $\Lambda (1405)$.  Especially a large decay width
$\Gamma _{\Lambda \gamma}$ directly indicates a large absolute value of
the $\bar{K}N$ compositeness $|X_{\bar{K}N}|$ and hence implies a large
$\bar{K}N$ component inside $\Lambda (1405)$.  On the other hand,
$\Sigma ^{0} \gamma$ decay is dominated by the $\pi \Sigma$ component
and hence the decay width becomes $\Gamma _{\Sigma ^{0} \gamma} \sim
23 \kev$ even for $|X_{\bar{K}N}|=0$.  Then, as $|X_{\bar{K}N}|$ grows
the values of $W^{\pm}$ are more separated from each other and the
allowed range for the $\Sigma ^{0} \gamma$ decay width is expanded.
Here we note that the maximal and minimal values of $\Gamma _{\Sigma
  ^{0}\gamma}$ become $\sim 40 \kev$ and $10 \kev$, respectively, for
$|X_{\bar{K}N}|=1$, so we could conclude that $|X_{\bar{K}N}|$ should
be large if the decay width for $\Sigma ^{0} \gamma$ would be
considerably large or considerably small.

By using the relation in Fig.~\ref{fig:rad_width_Lam} we can estimate
the $\bar{K}N$ compositeness from the $\Lambda (1405)$ radiative decay
width.  Actually, there are ``experimental'' data on the $\Lambda
(1405)$ radiative decay width evaluated from an isobar model fitting
of the decays of the $K^{-}p$ atom~\cite{Burkhardt:1991ms}: $\Gamma
_{\Lambda \gamma}=27 \pm 8 \kev$ and $\Gamma _{\Sigma ^{0} \gamma}=10
\pm 4 \kev$ or $23 \pm 7 \kev$.  From these ``experimental'' values we
can estimate the $\bar{K}N$ compositeness by using the relation in
Fig.~\ref{fig:rad_width_Lam}.  As a result, we extract
$|X_{\bar{K}N}|=0.5 \pm 0.2$ from $\Gamma _{\Lambda \gamma}=27 \pm 8
\kev$, $|X_{\bar{K}N}| > 0.5$ from $\Gamma _{\Sigma ^{0} \gamma}=10
\pm 4 \kev$, while $|X_{\bar{K}N}|$ can have an arbitrary value within
$\Gamma _{\Sigma ^{0} \gamma}=23 \pm 7 \kev$.  These results suggest
that the absolute value of the $\bar{K}N$ compositeness is
$|X_{\bar{K}N}|\gtrsim 0.5$, which implies that $\bar{K}N$ seems to be
the largest component inside $\Lambda (1405)$.

Finally we make several comments.  In this study we use the Particle
Data Group value to determine the pole position of $\Lambda (1405)$ as
$Z_{\rm pole}=M_{\Lambda (1405)} - i \Gamma _{\Lambda (1405)}/2$.
However, the $\Lambda (1405)$ pole position is not well determined,
although the compositeness~\eqref{eq:comp_KN-II} should be evaluated
on the $\Lambda (1405)$ pole position.  This may lead to an ambiguity
of the relation between the $\bar{K}N$ compositeness and the radiative
decay width shown in Fig.~\ref{fig:rad_width_Lam}.  This point is
discussed in the next subsection together with the effects of the
two-pole structure for $\Lambda (1405)$.

Besides, we have neglected the bare state contribution of $\Lambda
(1405)$ to the radiative decay.  Actually, even if the $\Lambda
(1405)$ would be dominated by a quark bound state such as $uds$ rather
than the meson-baryon component, $\Lambda (1405)$ would have finite
spatial size coming from the quark dynamics.  This would lead to the
additional contribution to the decay width, and hence the decay width
in Fig.~\ref{fig:rad_width_Lam} would be shifted upward.
Nevertheless, in this study we do not take into account such a
contribution since a usual constituent quark model cannot describe
$\Lambda (1405)$, which indicates that the ordinary quark
configuration inside $\Lambda (1405)$ is small.

We also note that our relation would be model dependent mainly from
the formulation of the radiative decay widths.  Actually we might
include form factors for the meson-baryon-baryon couplings, or we
might use a usual Dirac-field propagators for baryons.  These effects
altogether would lead to $\sim 10 \%$ errors.  Nevertheless, the
scenario that the larger radiative decay width to $\Lambda \gamma$
directly leads to the larger absolute value of the $\bar{K}N$
compositeness would not be changed.

\subsection{Analysis from the $\bm{\Lambda (1405)}$ pole position}
\label{sec:4c}

In the previous subsection we have obtained the relation between the
$\Lambda (1405)$ radiative decay width and the absolute value of the
$\bar{K}N$ compositeness with the $\Lambda (1405)$ pole position
determined from the value in Particle Data Group.  However, as we have
already mentioned, the $\Lambda (1405)$ pole position is not well
determined in experiments and moreover $\Lambda (1405)$ has two poles
according to the chiral unitary approach.  Therefore, in this
subsection we analyze how our relation between the radiative decay
width and the absolute value of the $\bar{K}N$ compositeness
$|X_{\bar{K}N}|$ depends on the $\Lambda (1405)$ pole position.

\begin{figure}[!t]
  \centering
  \Psfig{8.6cm}{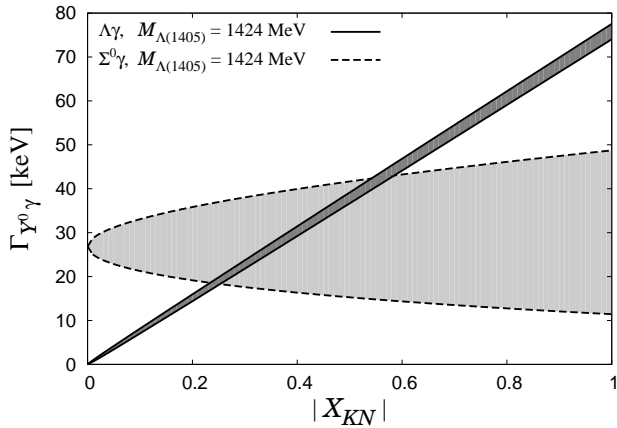}
  \caption{Radiative decay width of $\Lambda (1405)$ as a function of
    the absolute value of the $\bar{K}N$ compositeness.  The $\Lambda
    (1405)$ mass is fixed as $M_{\Lambda (1405)}=1424 \mev$.}
  \label{fig:rad_width_LamH}
\end{figure}

Firstly we show how the relation shown in Fig.~\ref{fig:rad_width_Lam}
is changed when the $\Lambda (1405)$ mass, {\it i.e.} the real part of
the pole position, shifts upward to $M_{\Lambda (1405)}=1424 \mev$, as
the higher $\Lambda (1405)$ pole in the chiral unitary approach.  In
this condition, nevertheless, we expect that the relation shown in
Fig.~\ref{fig:rad_width_Lam} will be not largely changed because we
have shown that for the $\bar{K}N$ bound state the binding energy
dependence of the relation between the $\bar{K}N$ compositeness and
the radiative decay width is very small (see
Fig.~\ref{fig:rad_width_BS}).  Indeed, by using $M_{\Lambda
  (1405)}=1424 \mev$ instead of $M_{\Lambda (1405)}=1405 \mev$, we
obtain the relation between the absolute value of the $\bar{K}N$
compositeness and the radiative decay width shown in
Fig.~\ref{fig:rad_width_LamH}.  The result with $M_{\Lambda
  (1405)}=1424 \mev$ is similar to that with $M_{\Lambda (1405)}=1405
\mev$ shown in Fig.~\ref{fig:rad_width_Lam} but the decay widths are
slightly larger according to the larger $\Lambda (1405)$ mass.  Then,
by using the relation in Fig.~\ref{fig:rad_width_LamH} we could
estimate the absolute value of the $\bar{K}N$ compositeness from the
``experimental'' value~\cite{Burkhardt:1991ms}: $|X_{\bar{K}N}|=0.4
_{-0.2}^{+0.1}$ from $\Gamma _{\Lambda \gamma}=27 \pm 8 \kev$,
$|X_{\bar{K}N}| > 0.6$ from $\Gamma _{\Sigma ^{0} \gamma}=10 \pm 4
\kev$, while $|X_{\bar{K}N}|$ can have an arbitrary value witnin $\Gamma
_{\Sigma ^{0} \gamma}=23 \pm 7 \kev$.  These results, especially from
$\Gamma _{\Lambda \gamma}$, would indicate that the absolute value of
the $\bar{K}N$ compositeness inside $\Lambda (1405)$ would decrease
slightly when a larger $\Lambda (1405)$ mass is used.

However, we should emphasize that the ``experimental'' value in
Ref.~\cite{Burkhardt:1991ms} is extracted from an isobar model fitting
of the decays of the $K^{-}p$ atom with the assumption $M_{\Lambda
  (1405)}=1405 \mev$.  Besides, in the chiral unitary approach the
$K^{-}p \to$ meson-baryon scatterings around and below the $K^{-}p$
threshold contains more weight on the higher $\Lambda (1405)$ pole of
the mass $\sim 1420 \mev$, which is indeed supported by the $K^{-}d
\to \pi \Sigma n$ reaction~\cite{Braun:1977wd, Jido:2009jf,
  Jido:2010rx, YamagataSekihara:2012yv, Jido:2012cy} and also by the
absorption branching ratios of $K^{-}$ from an atomic
orbit~\cite{Sekihara:2012wj}.  Therefore, it is better to make a
simple reanalysis of the data on the branching ratios $\Gamma _{K^{-}p
  \to \Lambda \gamma} / \Gamma _{K^{-}p \to \text{anything}}$ and
$\Gamma _{K^{-}p \to \Sigma ^{0} \gamma} / \Gamma _{K^{-}p \to
  \text{anything}}$ used in Ref.~\cite{Burkhardt:1991ms} with the mass
and the coupling constant $g_{K^{-}p}$ for the higher $\Lambda (1405)$
pole listed in Table~\ref{tab:2}.  Actually, by replacing the
parameters $M_{\Lambda (1405)}=1405 \mev$ and $g_{K^{-}p}=3.2$ used in
Ref.~\cite{Burkhardt:1991ms} with $M_{\Lambda (1405)}=1424 \mev$ and
$g_{K^{-}p}=2.25 + 0.87 i$, we have obtained the larger radiative
decay widths $\Gamma _{\Lambda \gamma} = 38 \pm 8 \kev$ and $\Gamma
_{\Sigma ^{0} \gamma} = 17 \pm 5 \kev$ or $42 \pm 7 \kev$ mainly due
to the larger $\Lambda (1405)$ mass.  Then, combined with the relation
in Fig.~\ref{fig:rad_width_LamH}, these values produce the absolute
value of the $\bar{K}N$ compositeness as: $|X_{\bar{K}N}|=0.5 \pm 0.1$
from $\Gamma _{\Lambda \gamma}=38 \pm 8 \kev$, $|X_{\bar{K}N}|>0.1$
from $\Gamma _{\Sigma ^{0} \gamma}=17 \pm 5 \kev$, and
$|X_{\bar{K}N}|>0.2$ from $\Gamma _{\Sigma ^{0} \gamma}=42 \pm 7
\kev$.  From this estimation, we can see that the result of the
absolute value of the $\bar{K}N$ compositeness from the $\Lambda
\gamma$ decay is consistent with that in the previous subsection using
$M_{\Lambda (1405)}=1405 \mev$ throughout the analysis.  This means
that our main conclusion of the absolute value of the $\bar{K}N$
compositeness from the $K^{-}p$ atom data will not be changed even if
the $\Lambda (1405)$ pole position is not well determined.  We note
that in order to reduce the ambiguity coming from analysis of the
experimental data, it is necessary to determine the $\Lambda (1405)$
radiative decay width in a model independent way.

\begin{figure}[!t]
  \centering
  \Psfig{8.6cm}{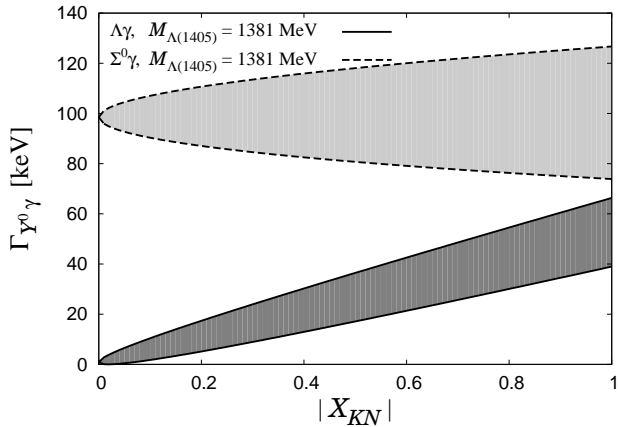}
  \caption{Radiative decay width of $\Lambda (1405)$ as a function of
    the absolute value of the $\bar{K}N$ compositeness.  The $\Lambda
    (1405)$ mass is fixed as $M_{\Lambda (1405)}=1381 \mev$ and the
    $\pi ^{\pm} \Sigma ^{\mp}$ coupling constants in Table~\ref{tab:2}
    are used.}
  \label{fig:rad_width_LamL}
\end{figure}

Next, in the two-pole scenario for $\Lambda (1405)$, the reaction
process controls which resonance pole has more weight.  Actually the
$\bar{K}N \to \pi \Sigma$ reaction process gives more weight to the
higher resonance pole around $1420 \mev$ while $\pi \Sigma \to \pi
\Sigma$ gives more weight to the lower pole~\cite{Jido:2003cb}.  For instance, in
Ref.~\cite{Geng:2007hz} the authors observe different shapes for the
$\Lambda \gamma$ and $\Sigma ^{0} \gamma$ invariant mass distributions
in the $K^{-}p \to \pi ^{0} Y^{0} \gamma$ and $\pi ^{-} p \to K^{0}
Y^{0} \gamma$ reactions, according to the different weight to the two
$\Lambda (1405)$ poles.  Namely, the former (latter) reaction is
dominated by the higher (lower) $\Lambda (1405)$ pole.  In the above
discussion on the radiative decay we have considered the higher pole
contribution.  Then, it is useful to discuss how the $\bar{K}N$
compositeness can be observed with the dominance of the lower pole
contribution such as in the $\pi ^{-} p \to K^{0} \Lambda \gamma$
reaction. Here we show in Fig.~\ref{fig:rad_width_LamL} the relation
between the absolute value of the $\bar{K}N$ compositeness inside the
lower $\Lambda (1405)$ and its radiative decay width with the $\Lambda
(1405)$ mass $M_{\Lambda (1405)}=1381 \mev$ and the $\pi ^{\pm} \Sigma
^{\mp}$ coupling constants listed in Table~\ref{tab:2}.  As one can
see, the branching ratios of the $\Lambda \gamma$ and $\Sigma ^{0}
\gamma$ decay modes are different from the previous cases; as a
reflection of the lower pole structure, the $\Sigma ^{0} \gamma$ is
dominant in the radiative decay even when $|X_{\bar{K}N}|$ is close to
unity.  We also observe a wider band both for the $\Lambda \gamma$
and $\Sigma ^{0} \gamma$ decay modes, which originates from the larger
$\pi \Sigma$ coupling constant $g_{\pi \Sigma}$.  Nevertheless, the
mean value of the allowed range for the $\Lambda \gamma$ decay shows
very similar behavior as that in Fig.~\ref{fig:rad_width_Lam}, which
means that we can extract similar absolute value of the $\bar{K}N$
compositeness with a certain value of $\Gamma _{\Lambda \gamma}$ from
both allowed regions of the $\Lambda \gamma$ mode in
Figs.~\ref{fig:rad_width_Lam} and \ref{fig:rad_width_LamL}.  From this
analysis, it is interesting to observe the $\Lambda (1405)$ radiative
decay in different $\Lambda (1405)$ production reactions, in which we
might observe the different branching ratios of the radiative decay as
an evidence of the two-pole structure for $\Lambda (1405)$ and also
the different $\bar{K}N$ compositeness for $\Lambda (1405)$.

\section{Conclusion}
\label{sec:5}

In this study we have investigated the radiative decay of $\Lambda
(1405)$, $\Lambda (1405) \to \Lambda \gamma$ and $\Sigma ^{0}
\gamma$, from the viewpoint of the $\bar{K}N$ compositeness, which
measures the amount of the $\bar{K}N$ component inside $\Lambda
(1405)$.  Since we can evaluate the $\bar{K}N$ compositeness by using
the $\Lambda (1405)$-$\bar{K}N$ coupling constant and the $\Lambda
(1405)$ pole position, we can establish a relation between the
radiative decay width and the $\bar{K}N$ compositeness by expressing
the radiative decay width with the $\bar{K}N$ coupling constant.

In order to grasp the behavior of the radiative decay width as a
function of the $\bar{K}N$ compositeness, we firstly consider a
$\bar{K}N(I=0)$ bound state without couplings to other channels.
Since there are one-to-one correspondences between the coupling
constant of the bound state to $\bar{K}N$ and the $\bar{K}N$
compositeness as well as the bound state-$\bar{K}N$ coupling constant
and the radiative decay width, we have established a relation between
the $\bar{K}N$ compositeness and the radiative decay width.
Especially the radiative decay width of the bound state is
proportional to the compositeness, since both the radiative decay
width and the compositeness are proportional to the squared bound
state-$\bar{K}N$ coupling constant.  We have obtained that the decay to
$\Sigma ^{0} \gamma$ is suppressed compared to the decay to $\Lambda
\gamma$ due to the strengths of the $K^{-}p\Lambda$ and $K^{-}p\Sigma
^{0}$ couplings.  Furthermore, we have found that the binding energy
dependence of the relation between the $\bar{K}N$ compositeness and
the radiative decay width is very small.

Bearing in mind the discussions on the radiative decay width of the
bound state, we have investigated the radiative decay of $\Lambda
(1405)$.  Here the absolute value of the $\Lambda (1405)$-$\bar{K}N$
coupling constant is determined from the absolute value of the
$\bar{K}N$ compositeness, while the $\Lambda (1405)$-$\pi \Sigma$
coupling constant is estimated from the strong decay of $\Lambda
(1405) \to \pi \Sigma$.  In order to take into account the
interference between $\bar{K}N$ and $\pi \Sigma$ in the decay
amplitude, we calculate the maximally constructive and destructive
interference, and we have shown the allowed region of the radiative
decay width of $\Lambda (1405)$ as a function of the absolute value of
the $\bar{K}N$ compositeness.  We note that the absolute value of the
$\bar{K}N$ compositeness cannot be interpreted as a probability of
finding the $\bar{K}N$ component but it will be an important piece of
information on the structure of $\Lambda (1405)$.  From the result of
the radiative decay width we have found that the allowed region for
the decay to $\Lambda \gamma$ is very narrow while that for the decay
to $\Sigma ^{0} \gamma$ is broad, since the decay to $\Lambda \gamma$
($\Sigma ^{0} \gamma$) is dominated by the $\bar{K}N$ ($\pi \Sigma$)
component inside $\Lambda (1405)$.  This means that the decay to
$\Lambda \gamma$ is suited to study the $\bar{K}N$ component in
$\Lambda (1405)$.  Furthermore, by using the ``experimental'' data on
the radiative decay width the absolute value of the $\bar{K}N$
compositeness is estimated as $|X_{\bar{K}N}|\gtrsim 0.5$, which
implies that $\bar{K}N$ seems to be the largest component inside
$\Lambda (1405)$.  We have also discussed the pole position dependence
of the radiative decay width, and have found that our main conclusion
of the absolute value of the $\bar{K}N$ compositeness from the
$K^{-}p$ atom data will not be changed even when the $\Lambda (1405)$
pole position is not well determined.  On the other hand, in the
two-pole scenario for $\Lambda (1405)$, we would observe the different
branching ratios of the radiative decay and the different $\bar{K}N$
compositeness for $\Lambda (1405)$ in different $\Lambda (1405)$
production reactions, which could be an evidence of the two-pole
structure for $\Lambda (1405)$.  Finally we emphasize that, in order
to evaluate more precisely the $\bar{K}N$ compositeness of $\Lambda
(1405)$ from experiments, it is necessary to determine precisely the
radiative decay width of $\Lambda (1405)$ in various production
reactions in a model independent way.

\begin{acknowledgments}

  This work was partially supported by the MEXT KAKENHI Grant Number
  25105010.

\end{acknowledgments}

\appendix

\section{Feynman rules}
\label{sec:A}

In this Appendix we summarize the Feynman rules used in this study.

In this study we use the following propagators
\begin{equation}
P_{m}( p^{2} ) = \frac{i}{p^{2} - m^{2} + i \epsilon} , 
\quad 
P_{M}( p^{2} ) = \frac{2 i M}{p^{2} - M^{2} + i \epsilon} , 
\end{equation}
for mesons and baryons, respectively, where $p$ is the momentum, $m$
and $M$ are the masses of the propagating meson and baryon,
respectively, and $\epsilon$ is an infinitesimal positive value.  The
Dirac matrices in $P_{M}$ are suppressed by an assumption that baryons
go almost on-shell.

Due to the requirement of the gauge invariance, the elementary
couplings of the photon to the mesons and baryons should be given by
the minimal coupling.  As a result, the $\gamma MM$ and $\gamma BB$
vertices are given as
\begin{equation}
\begin{split}
& - i V_{\gamma MM} = - i e Q_{M} \epsilon _{\mu} ( p + p^{\prime} )^{\mu} ,
\\ 
& - i V_{\gamma BB} = - i e Q_{B} \epsilon _{\mu} 
\frac{( p + p^{\prime} )^{\mu}}{2 M} ,
\end{split}
\end{equation}
with the elementary charge $e$, the charges of meson and baryon,
$Q_{M}$ and $Q_{B}$, respectively, the photon polarization $\epsilon
_{\mu}$, and the incoming and outgoing momenta for the hadrons
$p^{\mu}$ and $p^{\prime \mu}$, respectively.  Although the magnetic
moments of the baryons could contribute to the $\gamma BB$ vertex,
they are small and the contribution vanishes in the heavy baryon limit
in this study as discussed in Ref.~\cite{Geng:2007hz}.

The $MBB$ coupling can be obtained from the lowest-order $\SUN{3}$
chiral Lagrangian
\begin{equation}
{\cal L} = 
- \frac{D + F}{\sqrt{2} f} \trace [ \bar{B} \gamma ^{\mu} \gamma _{5} 
\partial _{\mu} \Phi B ] 
- \frac{D - F}{\sqrt{2} f} \trace [ \bar{B} \gamma ^{\mu} \gamma _{5} 
 B \partial _{\mu} \Phi ] , 
\end{equation}
with the meson decay constant $f$, parameters $D$ and $F$, and the
flavor $\SUN{3}$ matrices for the baryons $B$ and Nambu-Goldstone
bosons $\Phi$.  This Lagrangian generates the $MBB$ vertex as
\begin{equation}
- i V_{MBB} = - \tilde{V}_{MBB} \gamma ^{\mu} \gamma _{5} q_{\mu},
\end{equation}
\begin{equation}
\tilde{V}_{MBB} 
\equiv \alpha _{MBB} \frac{D + F}{2 f} 
+ \beta _{MBB} \frac{D - F}{2 f} , 
\end{equation}
with the incoming meson momentum $q^{\mu}$.  Then the nonrelativistic
reduction $\gamma ^{\mu} \gamma _{5} q_{\mu} \to - \bm{\sigma} \cdot
\bm{q}$ leads the vertex to 
\begin{equation}
- i V_{MBB} = \tilde{V}_{MBB} \bm{\sigma} \cdot \bm{q} 
= - \tilde{V}_{MBB} \sigma _{\mu} q^{\mu} ,
\end{equation}
with $\sigma ^{\mu} = (0, \, \bm{\sigma})$.

Finally the $\gamma MBB$ vertex is obtained by applying the minimal
coupling with respect to the $MBB$ coupling as
\begin{equation}
- i V_{\gamma MBB} 
= - e Q_{M} \bm{\epsilon} \cdot \bm{\sigma} \tilde{V}_{MBB}
= e Q_{M} \epsilon _{\mu} \sigma ^{\mu} 
\tilde{V}_{MBB}
\end{equation}
where the nonrelativistic reduction has been used.

\section{Ward identity for the radiative decay amplitude}
\label{sec:B}

\begin{figure}[!t]
  \centering
  \begin{tabular}{cc}
    \PsfigII{0.22}{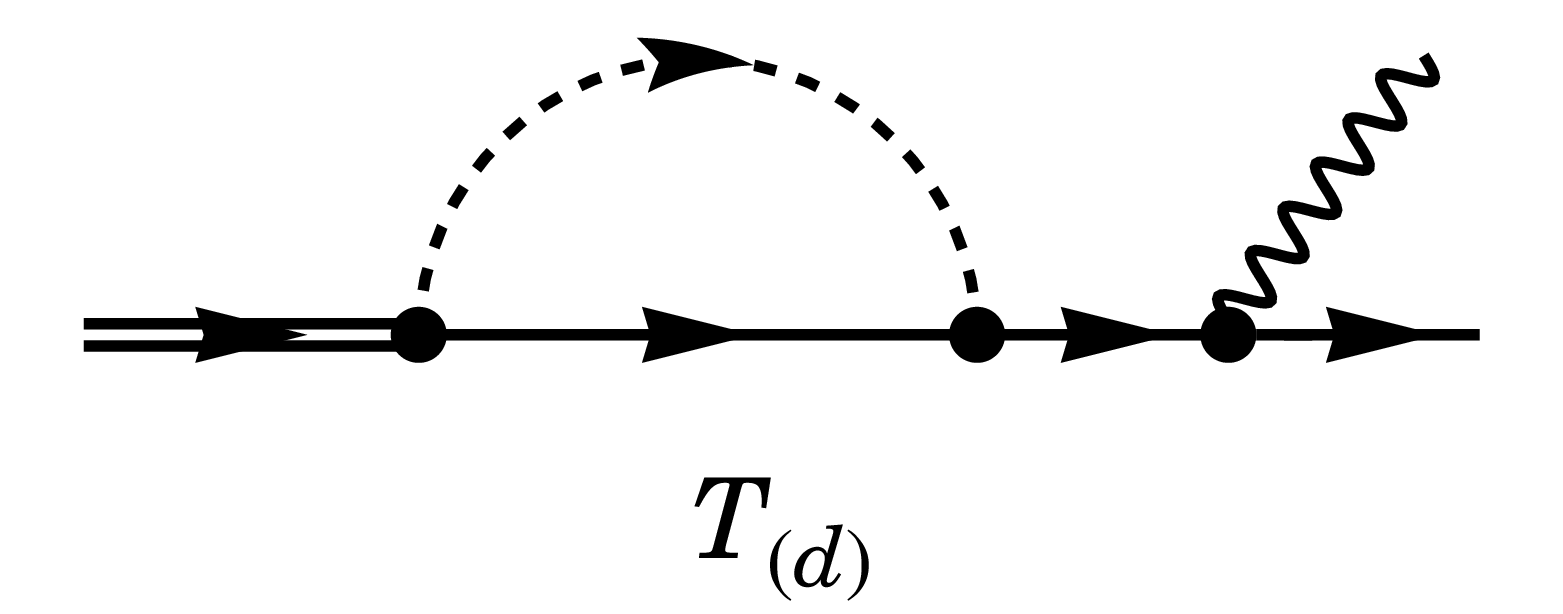} & 
    \PsfigII{0.22}{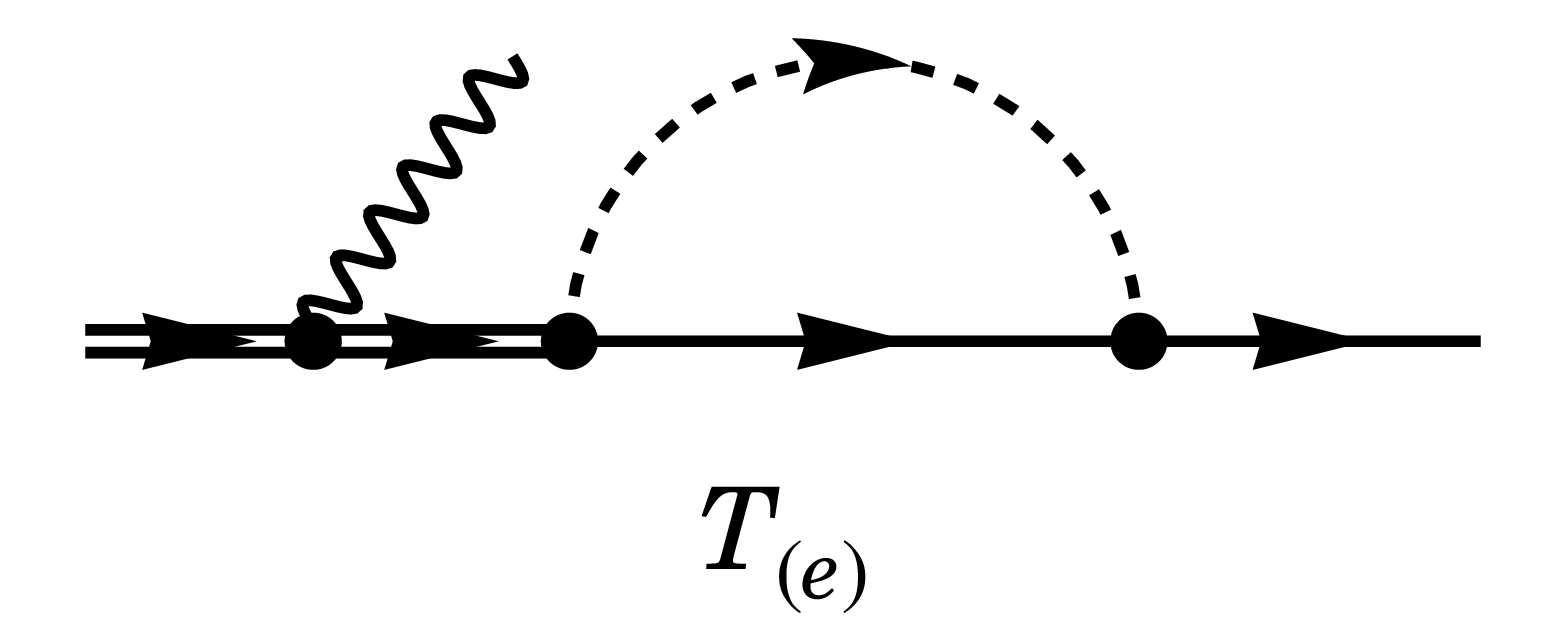}
  \end{tabular}
  \caption{Supplemental Feynman diagrams for the radiative
    decay with nonzero total charge. }
  \label{fig:diag_rad_2}
\end{figure}

In this Appendix we prove the Ward identity for the radiative decay
amplitude used in this study.  Here we consider a general case with
nonzero total charge $Q_{T}=Q_{M}+Q_{B}\ne 0$, which is not the case
of the $\Lambda (1405)$ radiative decay, but consider only the single
channel.  Due to the nonzero total charge, we have diagrams for the
radiative decay shown in Fig.~\ref{fig:diag_rad_2} in addition to them
in Fig.~\ref{fig:diag_rad}.  By using the Feynman rules and the
condition $k^{2}=0$, their decay amplitudes can be expressed as
\begin{equation}
- i T_{(d)} = + e g Q_{T} \tilde{V} \sigma _{\mu} \epsilon _{\nu}^{\ast} 
\frac{( 2 P - k )^{\nu}}{2 P \cdot k} L^{\mu} ( P ) ,
\end{equation}
\begin{equation}
- i T_{(e)} = - e g Q_{T} \tilde{V} \sigma _{\mu} \epsilon _{\nu}^{\ast} 
\frac{( 2 P - k )^{\nu}}{2 P \cdot k} L^{\mu} ( P - k ) ,
\end{equation}
where the loop integral $L^{\mu}(P)$ is defined as
\begin{equation}
L^{\mu} ( P ) \equiv i \int \frac{d ^{4} q}{(2 \pi)^{4}} 
\frac{q^{\mu}}{q^{2} - m^{2}} 
\frac{1}{(P - q)^{2} - M^{2}} , 
\end{equation}
Other three amplitudes are given in Eqs.~\eqref{eq:Ta}, \eqref{eq:Tb},
and \eqref{eq:Tc}.

The Ward identity states that sum of the five amplitudes becomes zero
if one takes $\epsilon _{\nu}^{\ast} \to k_{\nu}$ for on-shell photon,
$k^{2}=0$.  In order to prove this, we firstly show the relations for
$k_{\nu}D_{1}^{\mu \nu}$ and $k_{\nu}D_{2}^{\mu \nu}$:
\begin{equation}
k_{\nu} D_{1}^{\mu \nu} 
= L^{\mu} ( P - k ) - L^{\mu} ( q ) + k^{\mu} G ( P ) , 
\end{equation}
\begin{equation}
k_{\nu} D_{2}^{\mu \nu} 
= L^{\mu} ( P - k ) - L^{\mu} ( q ) , 
\end{equation}
which can be obtained by using the identity: 
\begin{equation}
\frac{k_{\nu} ( 2 q - k ) ^{\nu}}
{[(q - k)^{2} - m^{2}] [q^{2} - m^{2}]} 
= \frac{1}{(q - k)^{2} - m^{2}} - \frac{1}{q^{2} - m^{2}} , 
\end{equation}
with $k^{2}=0$.  Then, with the replacement of $\epsilon _{\nu}^{\ast}
\to k_{\nu}$, the decay amplitudes become 
\begin{equation}
T_{(a)} 
\to - i e g Q_{M} \tilde{V} \sigma _{\mu} k^{\mu} G(P) ,
\end{equation}
\begin{equation}
T_{(b)} 
\to + i e g Q_{M} \tilde{V} \sigma _{\mu} [ L^{\mu} ( P - k )
- L^{\mu} ( P ) + k^{\mu} G ( P ) ] ,
\end{equation}
\begin{equation}
T_{(c)} 
\to + i e g Q_{B} \tilde{V} \sigma _{\mu} [ L^{\mu} ( P - k )
- L^{\mu} ( P ) ] ,
\end{equation}
\begin{equation}
T_{(d)} 
\to + i e g Q_{T} \tilde{V} \sigma _{\mu} L^{\mu} ( P ) ,
\end{equation}
\begin{equation}
T_{(e)} 
\to - i e g Q_{T} \tilde{V} \sigma _{\mu} L^{\mu} ( P - k ) ,
\end{equation}
As a result, we obtain
\begin{equation}
T_{(a)} + T_{(b)} + T_{(c)} + T_{(d)} + T_{(e)} \to 0 . 
\label{eq:Ward}
\end{equation}
for $\epsilon _{\nu}^{\ast} \to k_{\nu}$.  This means that the Ward
identity is indeed satisfied for the five amplitudes.  Here we have
assumed the decay in the single channel, but Eq.~\eqref{eq:Ward}
indicates that the Ward identity is satisfied in each channel.  This
is reasonable, because in the multi-channels approach we may take the
coupling constant $g_{i}$ independently for each channel $i$ and hence
the Ward identity should be satisfied independently in each channel.
Finally we emphasize that if $Q_{T}=0$, as the case of the $\Lambda
(1405)$ radiative decay, we do not have $T_{(d)}$ and $T_{(e)}$ and
hence the Ward identity is satisfied even for the three amplitudes
$T_{(a)}$, $T_{(b)}$, and $T_{(c)}$.

\end{document}